\newif\ifreview

\ifreview
\documentclass[sigconf,natbib,anonymous,review]{acmart}
\else
\documentclass[sigconf,natbib]{acmart}
\fi

\renewcommand\footnotetextcopyrightpermission[1]{}

\AtBeginDocument{%
  \providecommand\BibTeX{{%
    \normalfont B\kern-0.5em{\scshape i\kern-0.25em b}\kern-0.8em\TeX}}}

\usepackage{rotating}

\usepackage[utf8]{inputenc}
\usepackage[T1]{fontenc}
\usepackage[nolist,printonlyused]{acronym}
\usepackage{float}
\usepackage{color,soul}
\sethlcolor{pink}
\usepackage{array}
\usepackage{multirow}
\usepackage{subcaption}
\usepackage{dirtytalk}
\usepackage{pifont}
\usepackage[group-minimum-digits=3]{siunitx}
\usepackage{colortbl}
\usepackage{tabularx}

\def\secref#1{\S\ref{#1}}
\def\figref#1{Fig.~\ref{#1}}
\def\tabref#1{Tab.~\ref{#1}}

\def\eg{\emph{e.g.}}
\def\ie{\emph{i.e.}}

\def\rq#1{\textbf{(RQ#1)}}
\def\c#1{\textbf{(C#1)}}

\def\makeanon#1{\ifreview[ANONYM]\else#1\fi}

\def\y{\ding{51}}

\setcopyright{none}
\acmBooktitle{Computer \& Security}

\begin{document}
\pagestyle{plain} %

\title%
    [\acs*{cvss} Prediction based on \acs*{osint} Information Sources]%
    {\acl*{cvss} Prediction based on \acl*{osint} Information Sources}

\begin{abstract}
The number of newly published vulnerabilities is constantly increasing. 
Until now, the information available when a new vulnerability is published is manually assessed by experts using a \acf*{cvss} vector and score.
This assessment is time consuming and requires expertise.
Various works already try to predict \acs*{cvss} vectors or scores using machine learning based on the textual descriptions of the vulnerability to enable faster assessment.
However, for this purpose, previous works only use the texts available in databases such as \acl*{nvd}.
With this work, the publicly available web pages referenced in the \acl*{nvd} are analyzed and made available as sources of texts through web scraping.
A \acl*{dl} based method for predicting the \acs*{cvss} vector is implemented and evaluated.
The present work provides a classification of the \acl*{nvd}'s reference texts based on the suitability and crawlability of their texts.
While we identified the overall influence of the additional texts is negligible, we outperformed the state-of-the-art with our \acl*{dl} prediction models.
\end{abstract}

\ifreview\else\author{Philipp Kühn}
\email{kuehn@peasec.tu-darmstadt.de}
\orcid{0000-0002-1739-876X}
\authornote{Corresponding author}

\author{David N. Relke}
\email{david.relke@stud.tu-darmstadt.de}

\author{Christian Reuter}
\email{reuter@peasec.tu-darmstadt.de}
\orcid{0000-0003-1920-038X}

\affiliation{%
  \institution{Science and Technology for Peace and Security (PEASEC), Technical University of Darmstadt}
  \city{Darmstadt}
  \country{Germany}}

\renewcommand{\shortauthors}{Kuehn~et~al.}\fi

\maketitle

\section{introduction}
\label{sec:introduction}
IT systems are now ubiquitous and fundamental to society, businesses, and individuals. 
Failures and disruptions can have catastrophic consequences for those affected. 
In 2017, for example, two waves of ransomware attacks occurred, each resulting in major outages to businesses and infrastructure~\citep{dwoskin_nations_2017, greenberg_untold_2018}. 
The vulnerability that enabled these attacks had been known and fixed a month before the first attack. 
In other attacks, such as the one on Microsoft Exchange Server in early 2021, only a few days passed between the discovery of the vulnerability and the start of attacks~\citep{brian_krebs_basic_2021}.

It is therefore important for researchers or system administrators to learn about vulnerabilities as early as possible, analyze them and initiate countermeasures. 
Various publicly accessible databases, such as the \ac{nvd}\footnote{\href{https://nvd.nist.gov/}{nvd.nist.gov}} and the \ac{cve}\footnote{\href{https://cve.mitre.org/}{cve.mitre.org}} collect, structure and prepare the published vulnerabilities for this purpose. 
However, relevant information can also be found on many other platforms, such as social media (especially Twitter), blogs, news portals, and company websites.

The \acf{cvss} is used to categorize different aspects of vulnerabilities. 
The result of this categorization is a vector whose elements are a machine-readable representation of the vulnerability's properties\footnote{\url{https://www.first.org/cvss/specification-document}}. 
Based on the components of the \ac{cvss} vector a numerical vulnerability score~(\ac{cvss} severity score) is calculated. 
The vulnerability assessment is usually performed by IT security experts based on the available \ac{osint} information. 
\ac{osint} refers to the structured collection and analysis of information that is freely available to the public.

There is a certain period of time when the information about a new vulnerability is published, but the assessment made by experts is not yet available~\citep{ruohonen_look_2019, elbaz_fighting_2020}.
Due to the large mass of published vulnerabilities, it is difficult for researchers or e.g. responsible persons in companies to assess each new vulnerability themselves.
They are therefore dependent on the assessments of experts.
Accordingly, the longer it takes for the assessment to become available, the longer it takes for countermeasures to be taken to mitigate the vulnerability.
During this period, the vulnerable systems are vulnerable to attack without the responsible parties knowing about it.
It is therefore important that the assessment is available as soon as possible.

Various works~\citep{elbaz_fighting_2020, han_learning_2017, shahid_cvss-bert_2021} try to perform this assessment automatically based on the textual information available about a vulnerability using \ac{ml}. 
This would allow for a much faster assessment. 
The vulnerability could already be assessed in an automated way when it is published and the time window in which no at least preliminary assessment is available is kept small.
It would also allow experts to prioritize and make recommendations for the assessment.

Previous work largely uses only the short descriptions of vulnerabilities from \ac{nvd} and \ac{cve} with some exceptions~\citep{chen_vest_2019, almukaynizi_proactive_2017}.
\citet{han_learning_2017}, for instance, present a system for classifying vulnerabilities into different severity levels based on \ac{cvss}.
From \citet{khazaei_automatic_2016} comes a work on predicting the numerical \ac{cvss} severity score.
In addition, there are methods that automatically predict the entire \ac{cvss} vector~\citep{elbaz_fighting_2020}.
Another work by \makeanon{\citet{kuehn_ovana_2021}} describes a system that uses Deep Learning to predict the \ac{cvss} vector.
However, the system requires labels created by experts to train, which significantly increases the required effort for larger datasets.
Further, \ac{dl} profits from large training datasets to which the reference texts could contribute, which is currently not leverage by related work.

\paragraph{Goal}
\label{par:goal}
This work aims to use as much textual data as possible to predict the \ac{cvss} vector of a vulnerability. 
This is to achieve the most accurate estimation of the \ac{cvss} vector possible.
It should be possible to use not only the short description of the vulnerability, but also other types of texts, such as Twitter posts and news articles for prediction in case of a new vulnerability.
Possible sources of textual information about vulnerabilities should be found and categorized. 
We aim to answer the following research questions:
\emph{Where can relevant textual information on vulnerabilities be found outside vulnerability databases~\rq{1}?} and
\emph{To which degree are public data sources beyond vulnerability databases suitable for predicting the \ac{cvss} vector~\rq{2}}?
This will clarify whether there are typical sources that regularly report on current vulnerabilities and whether these are suitable as a basis for building a dataset for training a \ac{ml} system.

Here, a first impression shall be gained by a rough manual search and then the sources referenced in the databases shall be analyzed automatically with regard to the type and scope of the references~(\eg, blog posts, patchnotes, GitHub issues). 
With the help of the texts, a \ac{ml} model for predicting the \ac{cvss} vector is to be trained.
The data must be filtered and cleaned for this purpose. 
The \ac{ml} model shall use \acl{dl} and use state-of-the-art models as a basis. 
The model is evaluated and compared to previous work.

\paragraph{Contributions}
\label{par:contributions}
The contribution to current research is an analysis of the references contained in the databases.
This will categorize the references in terms of certain characteristics and suitable for \ac{ml} models and can serve as a starting point for further work on the use of the references~\c{1}.
A method that collects and processes the text contained on the referenced web pages will be presented.
In addition, a system is implemented and evaluated that, unlike previous work, such as \citet{elbaz_fighting_2020} and \makeanon{\citet{kuehn_ovana_2021}}, uses more extensive text from the references in addition to descriptions of vulnerabilities from the databases~\c{2}.
This method for predicting \ac{cvss} vectors surpasses the current state-of-the-art.
Further, do we present an extensive explainability analysis of our trained models as part of our evaluation~\c{3}.

\paragraph{Outline}
\label{par:outline}
The state of the art in research is considered in \secref{sec:related_work}, followed by a preliminary analysis of the references included in \ac{nvd}~\see{\secref{sec:pre-analysis}}.
Requirements for references and the texts contained in them are defined and consequently the individual references are evaluated, resulting in a selection of references.
\secref{sec:implementation} explains the procedure for collecting the texts from the references and a system for retrieving, processing, and storing the texts is presented.
\secref{sec:evaluation} evaluates the \ac{ml} system, while \secref{sec:discussion} discusses and compares the results with other work.
Finally, a conclusion is drawn in \secref{sec:conclusion}.
\section{Related Work}
\label{sec:related_work}
This section gives an overview over the state of the art in research.
We focus literature dealing with the prediction of \ac{cvss} vectors, scores, or levels.
In addition, work that uses sources other than \ac{nvd} in this context is considered.
Automated assessment should provide a time advantage over the assessment by human experts.
In this regard, different papers come to different conclusions regarding the duration of the assessment, and the exact methodology is not always clear. 
\citet{elbaz_fighting_2020} state for the observed period from 2007 to 2019 that 90\% of vulnerabilities were assessed within just under 30 days, with a median of only one day, while \citet{chen_vest_2019} indicate an average of 132 days between publication and assessment for an observed period of 23 months in 2018 and 2019.

\paragraph{NVD, CVSS, Information Sources}
\citet{johnson_can_2018-1} perform a statistical analysis of \ac{cvss} vectors in different databases containing vulnerabilities.
In doing so, they show that despite different sources, the \ac{cvss} vector is always comparable and, consequently, seem to be robust.
They state the \ac{nvd} is the most robust information source for \ac{cvss} information.
On the other hand, \citet{dong_towards_2019} show that information in the \ac{nvd} itself is sometimes inconsistent and propose a system that relies on external sources to find, for example, missing versions of the software in question in the \ac{nvd}.
Accordingly, \makeanon{\citet{kuehn_ovana_2021}} present an information quality metric for vulnerability databases and improve several drawbacks in the \ac{nvd}.
In addition to vulnerability databases, other sources of information are used in vulnerability management.
\citet{sabottke_vulnerability_2015} use Twitter to predict whether a vulnerability will actually be exploited.
\citet{almukaynizi_proactive_2017} go a step further and use other data sources, such as ExploitDB\footnote{\url{https://www.exploit-db.com/}} and Zero Day Initiative\footnote{\url{https://www.zerodayinitiative.com/}}.
However, no text is used, but the simple existence of an article about a vulnerability is used as a feature for the \ac{ml} model.

\paragraph{CVSS Prediction}
A large number of works deal with the prediction of \ac{cvss} vector, scores, or levels starting from text.
As one of the first works, \citet{yamamoto_text-mining_2015} use sLDA \citep{mcauliffe_supervised_2007} to predict the \ac{cvss} vector based on the descriptions.
For predicting the score, \citet{khazaei_automatic_2016} use \acp{svm}, random forests~\citep{breiman_random_2001}, and fuzzy logic.
\citet{spanos_multi-target_2018} predict the \ac{cvss} vector using random forests and boosting~\citep{freund_short_1999}. 
\ac{dl} is first used in this context by \citet{han_learning_2017}.
By using an \ac{cnn}, no feature engineering is required.
However, in doing so, the model only determines the \ac{cvss} severity level from the options \textit{Critical}, \textit{High}, \textit{Medium}, and \textit{Low}.
\citet{gawron_automatic_2018} use \ac{dl} in addition to Naive Bayes, but here the result is a \ac{cvss} vector.
Twitter serves as the data source for \citet{chen_vase_2019-1}.
The \ac{ml} model is based on \ac{lstm}~\citep{hochreiter_long_1997} and predicts \ac{cvss} score.
\citet{sahin_conceptual_2019} also improve on the \citet{han_learning_2017} approach by using a \ac{lstm}.
\citet{gong_joint_2019} show a multi-task learning method that sets up multiple classifiers on a single \ac{nn}, making it more efficient.
\citet{liu_vulnerability_2019} use the Chinese equivalent, the \ac{cnnvd}, as the data source rather than the \ac{nvd}.
\citet{jiang_approach_2020} take scores not only from the \ac{nvd} but also from other sources as a basis for their prediction of the score.
The work of \citet{elbaz_fighting_2020} focuses on a particularly tractable classification of the \ac{cvss} vector.
Therefore, they do not use dimension reduction techniques.
\makeanon{\citet{kuehn_ovana_2021}} use \ac{dl} to predict the \ac{cvss} vector, based on the \ac{nvd}'s descriptions, with the goal to aid security experts in their final decision.
The most recent approach proposed \citet{shahid_cvss-bert_2021}, which uses a separate classifier based on a \ac{bert} model~\citep{devlin_bert_2019} to determine the \ac{cvss} vector for each component of the vector. 
Several proposals rely solely on the textual data from the \ac{nvd}.
Some use text from Twitter or simple binary features, such as the existence of an article about a particular vulnerability.
Other vulnerability context tasks also use few different data sources.
\citet{yitagesu_automatic_2021} also use Twitter as a source for a model for \ac{pos} tagging.
\citet{liao_acing_2016} propose a system which draws on several sources to filter \ac{ioc} from natural text.

\paragraph{Research Gap}
\label{par:research gap}
\Ac{osint} is widely used in IT security~\citep{pastor-galindo_not_2020,sabottke_vulnerability_2015,liao_acing_2016,chen_using_2019}.
Various works exist on the prediction of \ac{cvss} vectors based on descriptions.
However, as research shows, few \ac{osint} vulnerability sources are used~\citep{le_survey_2021}, especially in the context of \ac{cvss} score, level, or vector prediction, and if they are, very simple features from other sources are used~\citep{almukaynizi_proactive_2017}.
Furthermore, there is no systematic analysis of the suitablility of \ac{nvd} references for \ac{cvss} vector prediction approaches.

\section{Preliminary Analysis}
\label{sec:pre-analysis}
The authors performed an exploratory analysis of the available data, \ie, vulnerability descriptions and outgoing references from the \ac{nvd}, to identify data suitability criteria and requirements for the web scraping process.
Suitable in the sense of the present work are texts that describe a vulnerability and can be directly assigned to a vulnerability via the \ac{cve} identification number.
In the following we list some assumptions we considered.
\begin{itemize}
    \item Each text shall be uniquely assignable to one and only one vulnerability via the \ac{cve} identification number.
    Without this criterion a text could be used as a training example for two different permutations of one of the components of the \ac{cvss} vector.
    This makes it difficult for the \ac{ml} algorithm to identify the relevant properties of the vulnerability.
    The vulnerabilities covered in a text may be very different, so it does not make sense to use the same text for multiple vulnerabilities.
    It is even possible that only one vulnerability is described, although several with different target vectors are mentioned.
    
    \item The texts should not contain the target variable, \ie, the \ac{cvss} vector.
    Otherwise, the \ac{ml} model could predict the target parameter based on the variable present in the input, without any actual meaningful learning effect.
    
    \item There should be as little noise as possible.
    This ensures a high quality of the prediction.
    As stated in \secref{sec:related_work}, the data otherwise contain patterns that could negatively affect the \ac{ml} model.
\end{itemize}

Our secondary goal with this exploratory analysis is to identify \emph{where to find usable data}, \emph{assess the data quality} and \emph{how it can be used}.
Those questions correlate with our research questions~\see{\secref{par:goal}}.

\subsection{Descriptions in the \acs*{nvd}}
The first and most important starting point for finding texts about vulnerabilities is the \ac{nvd}.
We consider \ac{nvd} entries from 2016 to 2021, based on the introduction of the current \ac{cvss} standard version 3.
Entries without \ac{cvss} version 3 information are excluded.
This is the case for vulnerabilities in 2016, when \ac{cvss}v3 was still in the process of wide adoption, and in 2021, where the \ac{cvss}v3 vector was not yet available at the time the entries were retrieved.
In total, we collected \num{88979} entries.

\begin{figure}
 \centering
 \includegraphics[width=\columnwidth]{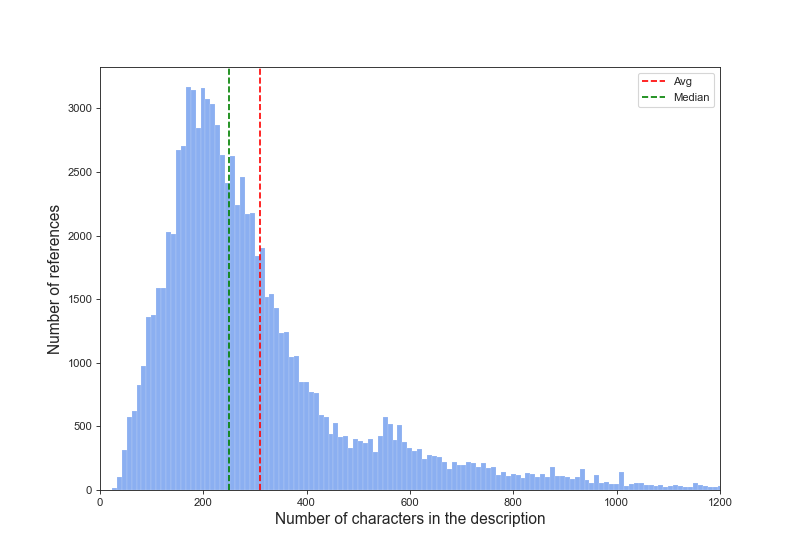}
 \caption{Distribution of \acl*{nvd} description lengths.}
 \label{fig:nvd_reference_length_dist}
\end{figure}

Individual entries in the \ac{nvd} contain a short, expert curated\footnote{\url{https://www.cve.org/ResourcesSupport/FAQs\#pc_cve_recordscve_record_descriptions_created}} description of the vulnerability.
The length of the descriptions for our collected entries ranges between \num{23} and \num{3835} characters, with an average of \num{310} and a median of \num{249}.
\figref{fig:nvd_reference_length_dist} shows the distribution of the length of the descriptions.
Descriptions longer than \num{1000} characters are very rare, with the 95$^{\mathit{th}}$ percentile already at \num{746} characters.
The information content of texts correlates with the pure length of the texts, apart from some exceptions\footnote{Some descriptions list other, non-identical, vulnerabilities, which artificially increases the length of the description without giving further content.}.
Likewise, a single, short sentence cannot describe all aspects of the vulnerability.
As \figref{fig:nvd_reference_length_dist} illustrates, there are a large number of vulnerabilities in \ac{nvd} with very short descriptions.

Literature shows that the quality of vulnerability descriptions in the \ac{nvd} differs~\makeanon{\citep{kuehn_ovana_2021}} and the quality can only be assessed to a limited extent without a deeper analysis.
A random sample shows that many descriptions contain less information about the actual vulnerability, but list, \eg, affected products and version numbers.
Such information is unrelated to the characteristics of the vulnerability and is therefore of little usefulness to predict the vulnerability severity.
Nevertheless, \citet{shahid_cvss-bert_2021} show that good results in the prediction of the \ac{cvss} vector are possible based only on \ac{nvd} descriptions.
Their method of \ac{cvss} score prediction achieves a \ac{mse} of \num{1.79} and a correctly predicted score in 53\% of all cases.

\subsection{Reference Analysis}
\label{subsec:analysis_references}
Each \ac{nvd} entry references websites.
To identify, which websites are suitable to be crawled we first analyze what kind of references are involved and, based on these insights, build categories for reference domains.
Second, we rate these groups based on their crawlability and potential text quality.

\begin{figure}
 \centering
 \includegraphics[width=\linewidth]{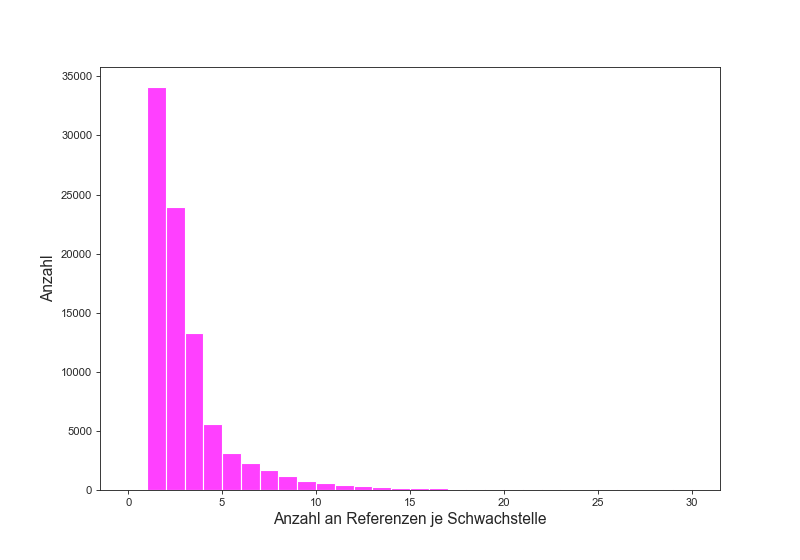}
 \caption{Distribution of number of references per vulnerability in \acl*{nvd}.}
 \label{fig:nvd_num_of_references}
\end{figure}

In the given subset of all entries of the \ac{nvd} there are a total of \num{251485} references.
The median number of references per vulnerability is \num{2}.
Many vulnerabilities have only a single reference, \num{95}\% have \num{8} or fewer references.
There are a few outliers with over \num{100} references.
The distribution of the number of references can be seen in \figref{fig:nvd_num_of_references}.

Over time, the diversity of references increased slightly.
From 2016 until 2021 there are \num{6013} different referenced domains, of which about \num{75}\% are accounted for by the \num{50} most frequent ones.
In 2016, \num{990} different domains are referenced with \num{86}\% of all references coming from the \num{50} most referenced websites.
In 2021, this trend increases to a total of \num{1711} domains referenced and the top \num{50} account for \num{74}\% of all references, showing an increase of diversity.
We build the \num{100} most frequently referenced \ac{nvd} reference domains based on our dataset~\see{\tabref{tab:nvd_most_frequent}}.
These domains account for \num{83}\% of all references in \ac{nvd}.

\tabref{tab:nvd_most_frequent} shows the \num{30} most frequently referenced domains, with additional entries to properly represent each group.

\begin{table}[]
    \centering
    \caption{The 30 most referenced domains in \ac{nvd}, with additional entries to properly represent each identified group. \textit{\#} gives the position in the Top-100, \textit{Num.} refers to how many times the given domain is referenced, \textit{Gr.} gives the assigned group, and \textit{Avail.} depicts, whether unavailable domain is unreachable, redirect to domains unrelated to the vulnerability, or are only reachable after login.}
    \begin{tabular}{@{}llrrr@{}}
    \toprule
    \textbf{\#} & \textbf{\acs{url}} & \textbf{Num.} & \textbf{Gr.}  &  \textbf{Avail.} \\ \addlinespace
    1 & github.com                     &  25064            &  1               &  \y       \\
    2 & www.securityfocus.com          &  20645            &  2               &  \y       \\
    3 & www.securitytracker.com        &  10842            &  2               &  \y       \\
    4 & access.redhat.com              &  8627             &  3/4             &  \y       \\
    5 & support.apple.com              &  8069             &  3               &  \y       \\
    6 & lists.opensuse.org             &  7930             &  2               &  \y       \\
    7 & lists.fedoraproject.org        &  7212             &  2               &  \y       \\
    8 & www.oracle.com                 &  7006             &  3               &  \y       \\
    9 & lists.apache.org               &  6294             &  2               &  \y       \\
    10 & www.debian.org                 &  5614             &  2/3             &  \y       \\
    11 & security.gentoo.org            &  5289             &  4               &  \y       \\
    12 & usn.ubuntu.com                 &  5225             &  3               &  \y       \\
    13 & lists.debian.org               &  4921             &  2               &  \y       \\
    14 & portal.msrc.microsoft.com      &  4391             &  3               &  \y       \\
    15 & www.openwall.com               &  4136             &  2               &  \y       \\
    16 & packetstormsecurity.com        &  4068             &  4               &  \y       \\
    17 & source.android.com             &  3672             &  3               &  \y       \\
    18 & seclists.org                   &  3462             &  2               &  \y       \\
    19 & www.exploit-db.com             &  3412             &  5               &  \y       \\
    20 & tools.cisco.com                &  3019             &  4/5             &  \y       \\
    21 & security.netapp.com            &  2890             &  5               &  \y       \\
    22 & www.ibm.com                    &  2807             &  4               &  \y       \\
    23 & exchange.xforce.ibmcloud.com   &  2673             &  4               &  \y       \\
    24 & helpx.adobe.com                &  2643             &  3               &  \y       \\
    25 & www.zerodayinitiative.com      &  2547             &  5               &  \y       \\
    26 & bugzilla.redhat.com            &  2482             &  1               &  \y       \\
    27 & rhn.redhat.com                 &  2019             &  4               &  \y       \\
    28 & www.mozilla.org                &  1785             &  4               &  \y       \\
    29 & crbug.com                      &  1458             &  1               &  \y       \\
    30 & www.ubuntu.com                 &  1397             &  3               &  \y       \\
    \multicolumn{4}{c}{\dots} \\
    33 & bugzilla.mozilla.org           &  1075             &  1               &  \y       \\
    \multicolumn{4}{c}{\dots} \\
    47 & wpscan.com                     &  791              &  5               &  \y       \\
    \multicolumn{4}{c}{\dots} \\
    66 & medium.com                     &  443              &  6               &  \y       \\
    \bottomrule
    \end{tabular}
    \label{tab:nvd_most_frequent}
\end{table}

We analyze which domains contain suitable descriptions of vulnerabilities and to what extent they are usable.
Based on their characteristics, we derive groups of references.
In the following, we present and describe the six identified groups in conjunction with some sample domains.

\begin{enumerate}

    \item \textbf{Version control and bug tracker services} \newline
    \textit{Examples: GitHub, crbug.com, bugzilla.mozilla.com} \newline
    These sites mostly contain program code, output and log files, technical descriptions, and bug discussions.
    A more abstract description of the vulnerabilities is rarely found.
    The domains are operated by the producers of the software, but contributions by users are also possible.
    Hence, there is not always an information verification by experts.
    On some sites the structure of the references is always identical, on others the structure is inconsistent.

    \item \textbf{Mailing Lists} \newline
    \textit{Examples: lists.fedoraproject.org, lists.apache.org, lists.debian} \newline
    Contributions origin from different individual users and are mostly unstructured and inconsistent texts and code fragments.
    As a result, some references to a domain may allow a unique mapping from \acs{cve}-ID to text, while this is not possible for other references to the same domain.
    Descriptions of vulnerabilities may be present, however, these are predominantly technical details.
    On some domains, vulnerabilities fixed with an update are also only mentioned without further text.

    \item \textbf{Patchnotes} \newline
    \textit{Examples: support.apple.com, oracle.com, helpx.adobe.com} \newline
    These are often maintained large commercial vendors.
    A single reference to one of the domains in this group typically contains information about many different vulnerabilities that have been closed with an update.
    On some domains, descriptions of the vulnerabilities are published, on others, the \ac{cve}-ID is only mentioned.
    References to one and the same domain have mostly identical structures over the whole observed period.
    The articles are written by employees of the respective companies.

    \item \textbf{Security Advisories} \newline
    \textit{Examples: tools.cisco.com, security.gentoo.org, ibm.com} \newline
    Vendors describe vulnerabilities in their own products in more detail on domains in this group.
    Often, only one vulnerability is covered in a reference.
    The structures of the references on a domain are the same.
    The descriptions of vulnerabilities are relatively detailed.
    The authors are employees of the respective companies.

    \item \textbf{Third party articles about vulnerabilities} \newline
    \textit{Examples: wpscan.com, zerodayinitiative.com, packetstormsecurity.com} \newline
    Companies or users publish articles on domains of this group about weak points in the products of other manufacturers.
    In some cases, this is part of a commercial business model based on services.
    Unlike the vulnerability-focused mailing lists, the structure of these posts is consistent.
    The contributions on some sites origin from professional employees, while on other sites unverified users are the authors of the texts.

    \item \textbf{Blog posts and social media} \newline
    \textit{Examples: medium.com, twitter.com, groups.google.com} \newline
    References to domains from this group show high diversity.
    The structure of the contributions is inconsistent.
    Authors may be professional contributors as well as unverified users.
    A clear assignment of \ac{cve}-ID to text depends on the authors of the specific contributions, not on the website itself.
    
\end{enumerate}

Our criteria for the suitability of texts for the training process of our \ac{ml} models~\see{\secref{sec:pre-analysis}} cannot be met by general purpose crawling approaches, like trafilatura~\citep{barbaresi_trafilatura_2021-1}, which ignore the characteristics of the target-domain.
Instead, solutions must be tailored to the target domain.
This is the only way to extract texts from the references that meet our requirements.
Since a large number of different domains are referenced, a pre-selection must be made.

The presented groups differ in terms of the usability of the references.
Within the groups the domains are differently suitable.
Ideal references allow a unique mapping from an \ac{cve}-ID to text.
The text must be an abstract description, since technical details such as code descriptions out-of-scope in the present work.
Since web scraper use the \ac{html} source code's structure of the domain to extract the correct text, individual references to a domain should therefore always have the same structure.

A unified structure is used on domains where contributions are published or at least reviewed by a single entity.
For the first and second group, there is only a higher-level structure, but not a uniform structure of the actual contribution.
For example, the basic structure of a reference to an issue in GitHub is always the same, however, the structure of the actual issue description might differ in each case.
\tabref{tab:usability_websites} shows an simplified overview of the different groups, whether they meet the uniqueness-, uniformity-, and abstract-text-requirements based on a 5-point scale.

\begin{table}
\caption{Overview of the 5-point scale evaluation for the usability of the different groups in combination with the usual origin of the content.}
\label{tab:usability_websites}
\centering
\begin{tabular}{@{}lllll@{}}
\toprule
Group & Origin                  & Unique     & Uniform    & Abs. text     \\ \addlinespace
VCS/Bug Tracker     & \textit{User}           & \y\y\y     & \y\y\y     & \y\y          \\
Mailing Lists     & \textit{User}           & \y\y       & \y         & \y\y          \\
Patchnotes     & \textit{Vendor}         & \y\y\y     & \y\y\y\y\y & \y\y          \\
Advisories     & \textit{Vendor}         & \y\y\y\y\y & \y\y\y\y\y & \y\y\y\y      \\
Third Party     & \textit{3$^\text{rd}$-P.} & \y\y\y\y   & \y\y\y\y\y & \y\y\y\y      \\
Blogs/Social Media     & \textit{User}           & \y\y       & \y         & \y\y\y        \\
\bottomrule
\end{tabular}
\end{table}

\paragraph{Domain Selection}
\label{par:domainselection}
For the domain selection, it must be considered whether it is worth the effort to adapt a web scraper for a domain.
Pages with the same structure and content require less effort and promise a better yield, as the texts will be more likely to meet the established criteria.

Starting from the frequency ranking of domains~\see{\tabref{tab:nvd_most_frequent}}, a domain selection is made based on the domains group and the group ranking of \tabref{tab:usability_websites}.

\begin{itemize}

    \item \textbf{\href{https://www.ibm.com/support/pages/node/6427953}{ibm.com}} \newline
    \textit{Group 4} - \num{3447} References \newline
    IBM publishes collected information about vulnerabilities in its own products.
    The individual texts are rather short.
    The assignment of text to \ac{cve} ID is easy thanks to the uniform structure of the articles.

    \item \textbf{\href{https://tools.cisco.com/security/center/}{tools.cisco.com}} \newline
    \textit{Group 4} - \num{3019} references. \newline
    Cisco publishes detailed descriptions for vulnerabilities in its own products or in third-party products that Cisco uses or integrates into its own products, such as frameworks.
    In addition, technical details and code are sometimes included.
    The structure of the articles is very similar.

    \item \textbf{\href{https://www.zerodayinitiative.com/about/}{zerodayinitiative.com}} \newline
    \textit{Group 5} - \num{2899} references. \newline
    Trend Micro\footnote{website: \href{https://www.trendmicro.com/de_de/business.html}{trendmicro.com/de\_en/business.html}} acts as a middleman between the discoverers of zero-day vulnerabilities and the manufacturers of the affected products.
    The advisories are then published.
    The structure and type of description are always the same.

    \item \textbf{\href{https://talosintelligence.com/vulnerability_reports/}{talosintelligence.com}} \newline
    \textit{Group 5} - \num{1335} References \newline
    Talos is a commercial company belonging to \href{https://www.cisco.com/}{Cisco} offering services and products related to IT security.
    The website publishes articles about vulnerabilities discovered by Talos.
    The articles are very detailed.
    The text on the website includes code, version numbers, \ac{cvss} vector and other information in addition to the description.
    However, the text itself is structured by headings that are consistent for all posts.

    \item \textbf{\href{https://www.qualcomm.com/company/product-security/bulletins/}{qualcomm.com}} \newline
    \textit{Group 3/4} - \num{1048} References \newline
    Contains information collected monthly on vulnerabilities in Qualcomm products.
    Descriptions are brief.
    The structure is consistent, and the articles are sorted into tables.
    Partially the \acsp{url} deposited in the \ac{nvd} are incorrect, because Qualcomm has changed the \ac{url} scheme over time. However, the monthly posts are still accessible under a modified \ac{url}.

    \item \textbf{\href{https://support.f5.com/csp/home}{support.f5.com}} \newline
    \textit{Group 5} - 932 references \newline
    F5 provides commercial IT security services and products.
    The referenced papers describe individual vulnerabilities in products developed by F5.
    The structure is consistent.

    \item \textbf{\href{https://wpscan.com/}{wpscan.com}} \newline
    \textit{Group 5} - 803 references. \newline
    A provider that rehashes vulnerabilities from the WordPress ecosystem and offers services related to the security of WordPress installations.
    For each \ac{cve} Identification number exists a short description, the structure of the page is the same throughout.

    \item \textbf{\href{https://www.intel.com/content/www/us/en/security-center/default.html}{intel.com}} \newline
    \textit{Group 4} - 771 references \newline
    Intel publishes here lists of vulnerabilities that have been fixed with an update.
    The structure of the pages is always identical and an assignment is possible without any problems.

    \item \textbf{\href{https://security.snyk.io/vuln/}{snyk.io}} \newline
    \textit{Group 5} - 671 references \newline
    Snyk offers several commercial vulnerability management products.
    The company maintains a public database of vulnerabilities in \ac{oss}, respectively in open source ecosystems like \ac{npm} or Maven.
    The descriptions are sometimes very detailed and the structure of the contributions is always identical.
\end{itemize}

The selected web pages are referenced a total of \num{14925} times.
However, it is to be expected that not all references are available anymore.

\paragraph{Special Features of Twitter}
Twitter is an important medium in IT security and has been the subject of several works~\citep{chen_using_2019,sabottke_vulnerability_2015}.
Twitter is also frequently referenced in \ac{nvd} and is found among the \num{100} most referenced websites.
However, a preliminary analysis shows that the references are unusable.
In some cases, only user profiles are referenced, such as for CVE-2021-25179\footnote{\url{https://nvd.nist.gov/vuln/detail/CVE-2021-25179}}.
The reference \url{twitter.com/gm4tr1x} is the profile of the vulnerability's discoverer\footnote{The \href{https://documentation.solarwinds.com/en/success_center/servu/content/release_notes/servu_15-2_release_notes.htm}{reference} to the SolarWinds vendor page lists the name \emph{Gabriele Gristina} as the discoverer. His \href{https://www.linkedin.com/in/gabrielegristina/\#experience}{LinkedIn} and \href{https://github.com/matrix}{GitHub} account are also referenced, in addition to the Twitter profile.}.
User profiles provide no meaningful information for the present work.
Generally, such references are not in line with the \ac{cve}'s reference requirements\footnote{\url{https://www.cve.org/ResourcesSupport/AllResources/CNARules\#section\_8-3\_cve\_record\_reference\_requirements}}.
In our dataset 17\% of references on Twitter are links to profiles.
Other references are retweets, such as seen in \href{https://nvd.nist.gov/vuln/detail/CVE-2021-27549}{CVE-2021-27549}\footnote{Referencing \url{https://twitter.com/0xabc0/status/1363855602477387783}}, yielding the same problem.
The original \href{https://twitter.com/0xabc0/status/1363788023956185090}{tweet} is also referenced in the \ac{nvd}.

Some Twitter references actually contain a description of the vulnerability.
Twitter is thus very important as a medium to exchange information between experts in a short amount of time, but cannot serve well as a source for texts in this work.

\section{implementation}
\label{sec:implementation}
While the previous section~\see{\secref{sec:pre-analysis}} examined the space of available references and accompanying requirements, this section explains the process of web scraping and model training.

\subsection{Web Scraping}
\label{subsec:implementation_web_scraping}
The selected domains~\see{\secref{par:domainselection}} are publicly available, but no \ac{api} exists to retrieve their content.
So the texts have to be extracted from the pages via web scraping.

Through the robots.txt\footnote{\url{https://www.robotstxt.org/}}, the operator of a website can select which bots should access which \acp{url}.
However, this employs only a soft restriction, since it cannot be technically enforced.
With the Python library \verb|urllib|, the robots.txt of the selected domain is checked whether access to the \ac{nvd} referenced in the \acp{url} is allowed.
In some cases, a delay between requests is desired due to the non-standard directive \textit{crawl-delay}.
The developed web scrapers respect this accordingly.

\begin{figure}[]
 \centering
 \includegraphics[width=\columnwidth]{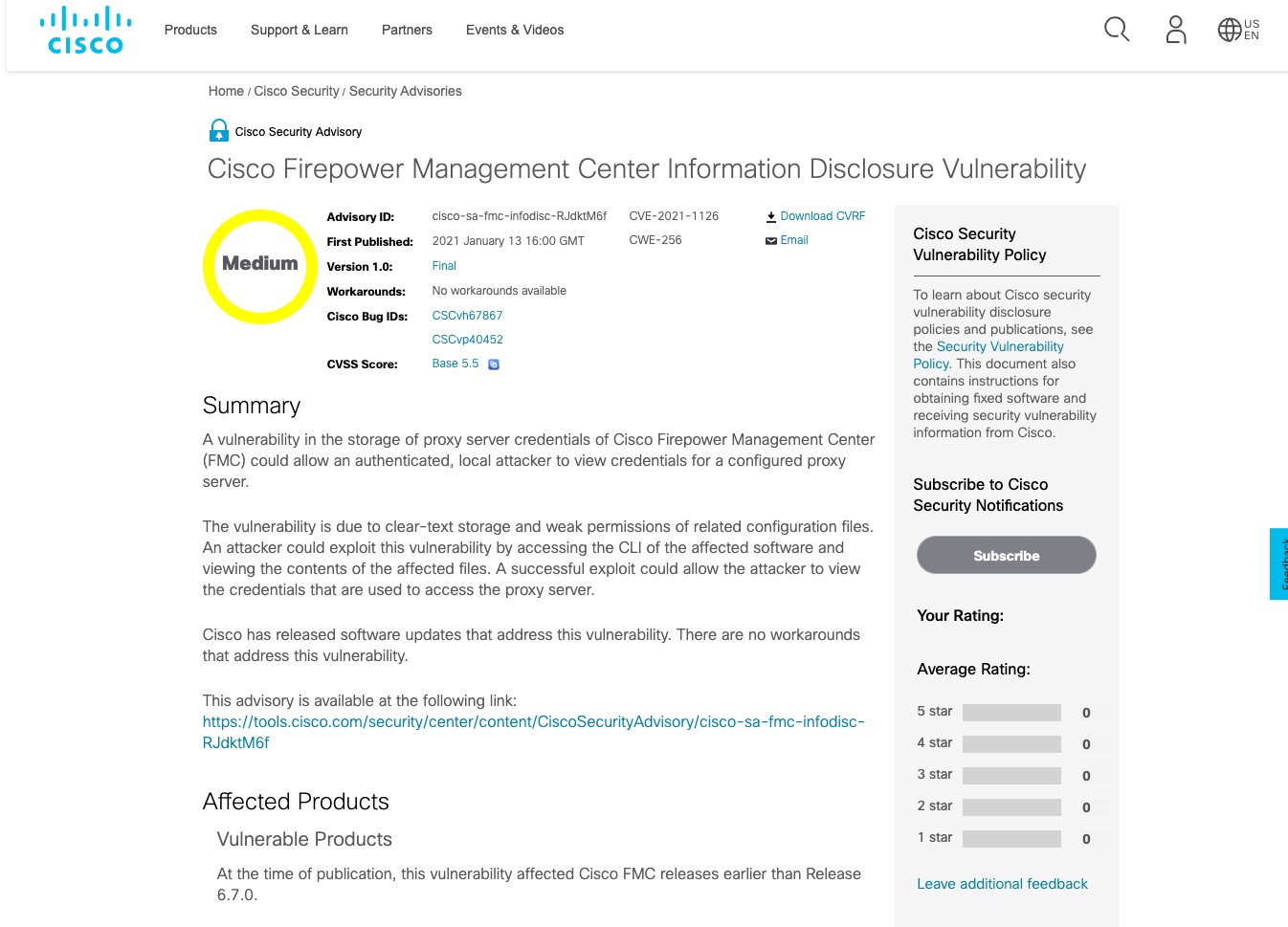}
 \caption{Example of Cisco website, referenced in \url{https://nvd.nist.gov/vuln/detail/CVE-2021-1148}.}
 \label{fig:cisco_screenshot}
\end{figure}

While trafilatura~\citep{barbaresi_trafilatura_2021-1} seem promising, our insights from \secref{sec:pre-analysis} show, that it should be avoided in the present work.
\figref{fig:cisco_screenshot} shows an example of the relevant part of Cisco's website.
It contains the requested description as well as other texts that is present on this page.
The static texts, such as headings and various legal information, are the same for each reference and represent noise.
While trafilatura removes parts such as the page header, bigger chucks like the legal information are still present during text extraction.

The results are similar for \textit{ibm.com}, \textit{zerodayinitiative.com}, \textit{wpscan.com}, \textit{talosintelligence.com}, and \textit{snyk.io}.
Some unwanted content could still be removed by filtering the output by trafilatura, but this would require post-processing, which negates the idea of trafilatura.
On some pages of \textit{qualcomm.com} and \textit{intel.com} multiple vulnerabilities are treated together, which introduces noise in the training process.
During implementation we identified, that, \eg, \textit{qualcomm.com} changed its \ac{url} structure, so that some of the referenced \acp{url} are unavailable.
However, a manual search shows that the pages themselves are still present under other \acp{url}.

Since Trafilatura cannot execute JavaScript, the pages of \textit{support.f5.com} cannot be retrieved at all.
This is because the server responds to initial \ac{http} GET requests for the referenced \acp{url} with a JavaScript file embedded in \ac{html}.
In a browser, the script is then executed and thus the actual page content is loaded.
While Trafilatura might work in other contexts, it is, in many ways, not suitable for the present work, partly due to the special requirements~\see{\secref{sec:pre-analysis}}.

Several other technologies offer better controllability and in-depth filtering capabilities.
Selenium\footnote{\url{https://www.selenium.dev/}} is a framework for automated testing of web applications and enables automatic control of full-featured web browsers in the background, \eg, Google Chrome and Mozilla Firefox.
Through \acp{api} for various programming languages, including Python, the web browser can be controlled.
The \acp{api} allow access to the \ac{dom} representation of the \ac{html} content of the accessed web page.
For testing, user interaction can be simulated, such as clicks or input.
Selenium thus provides everything necessary to JavaScript enriched web pages.
However, it is not a lightweight and particularly fast solution.

\textit{Beautiful Soup}\footnote{\url{https://www.crummy.com/software/BeautifulSoup/}} is an \ac{oss} web scraping library for Python.
It allows parsing of \ac{html} files.
The user can navigate through the \ac{api} structure to get selected parts of the web page.
Beautiful Soup is lightweight and faster than Selenium, but is limited to \ac{html} content.
If parts of the page are reloaded using JavaScript, Beautiful Soup cannot access them accordingly.

Since the amount of references to be retrieved with the Web Scraper is limited to \num{14925} and the retrieval is done only once, time plays only a minor role. 
Rendering the web pages with Selenium takes most of the time.
The speed can be increased linearly by parallelization.

The program is structured according to the producer-consumer design pattern.
First, all \acp{url} are collected, then multiple threads are started to process the \acp{url} in parallel.
The correct web scraper is selected based on the \ac{url}.

The web scrapers for \textit{talosintelligence.com} and \textit{intel.com} are implemented using Beautiful Soup, and Selenium is used for the rest of the pages.
The Beautiful Soup based web scrapers take about a second to retrieve and parse a web page, while Selenium based web scraper usually takes about five seconds.
The web scraper first waits until the requested page is fully loaded and no more JavaScript is executed.
Sometimes this leads to a blockade, because JavaScript is executed permanently.
Therefore, the execution is automatically interrupted after 20 seconds.
The page with the actual text is usually fully loaded by that time and can be parsed.
Since such timeouts occur seldom, resulting idle times are negligible.
In total, a complete run over all references in the selection took about 12 hours at a measured Internet speed of about 50~MBit/s and five parallel web scrapers.

As mentioned before, some \acp{url} for \textit{qualcomm.com} are unavailable.
Hence, the web scraper is set up to first check the \ac{nvd}'s reference and if it fails, start another attempt corrected \ac{url}, corresponding to the current \ac{url} scheme.
This reliably fixes the \ac{url} problems for \textit{qualcomm.com}.

\begin{table}
\centering
\caption{number and proportion of each reference successfully retrieved by the web scraper from the preselection}
\label{tab:successful_references}
\begin{tabular}{lSS[table-format=1.2]}
\toprule
\multicolumn{1}{c}{} & \multicolumn{2}{c}{References} \\ 
\cmidrule{2-3}
Webpage & {Crawled} & {Ratio} \\
\midrule
ibm.com                &  2868  & 0.83   \\
tools.cisco.com        &  3004  & 0.99   \\
zerodayinitiative.com  &  2899  & {\cellcolor{green}} 1.0    \\
talosintelligence.com  &  1201  & 0.89   \\
qualcomm.com           &   697  & 0.66   \\
support.f5.com         &   740  & 0.79   \\
wpscan.com             &    35  & {\cellcolor{red!50}} 0.04   \\
intel.com              &   731  & 0.94   \\
snyk.io                &   627  & 0.93   \\ \addlinespace
Total                  & 12802  & 0.85   \\
\bottomrule
\end{tabular}
\end{table}

In total, \num{12755} references~(85\%) of the \num{14925} original ones are successfully retrieved.
During the crawling process, we identified problems with \textit{wpscan.com}.
The domain permits all bot access in its robots.txt, but blocks all requests after five initial ones in quick succession.
This means that it is not possible to retrieve a large number of references in a meaningful way.
Of the \num{803} references originally available, only \num{35} were retrieved.

\begin{figure}[]
 \centering
 \includegraphics[width=\columnwidth]{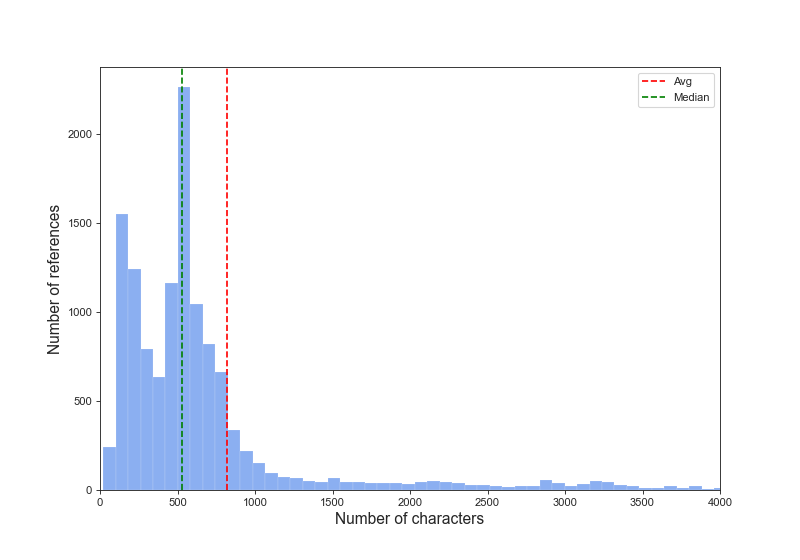}
 \caption{Distribution of text lengths retrieved with the web scraper.}
 \label{fig:dist_scraper_text_length}
\end{figure}

\figref{fig:dist_scraper_text_length} shows the distribution of the lengths of the successfully retrieved texts.
The average is \num{817.11}, and the median is \num{532} characters.
The texts are between \num{32} and \num{32206} characters long.
Thus, the obtained texts are significantly longer than the descriptions from \ac{nvd}~\see{\figref{fig:nvd_reference_length_dist}}.

\figref{fig:dist_scraper_text_length} shows the distribution of lengths of successfully retrieved texts.
The average is \num{817.11}, the median is 532 characters.
The texts are between 32 and \num{32206} characters long.
Thus, the obtained texts are significantly longer than the descriptions from \ac{nvd}~\see{\figref{fig:nvd_reference_length_dist}}.

\subsection{\acl*{dl} Classifier}
\label{subsec:dl_classifier}
The goal of the work is to predict the entire \ac{cvss} basis vector, the problem is split into several subproblems in the form of classifying the individual components of the vector.
The components of the \ac{cvss} vector are \acf{av}, \acf{ac}, \acf{pr}, \acf{ui}, \acf{s}, \acf{c}, \acf{i}, and \acf{a}\footnote{\url{https://www.first.org/cvss/specification-document}}.
For each component there is an independent classifier.
As a result, eight models must be trained separately.

\subsubsection{Model Selection}
\label{subsubsec:selection_models}
\citet{shahid_cvss-bert_2021} use a model based on \ac{bert}~\citep{devlin_bert_2019} for their work.
A classifier in the form of a fully-connected feed-forward \ac{nn} is placed on top of the \ac{bert} base model in each case.
\citet{shahid_cvss-bert_2021} use \acs{bert}-small~\citep{turc_well-read_2019}, rather than the original version of \ac{bert}~\citep{devlin_bert_2019}.
This model achieves a similar result in various benchmarks with significantly fewer parameters than \ac{bert}, but is faster to train.
Since eight models must be trained, we adapt this idea to use one of the smaller \ac{bert} models.
DistilBERT~\citep{sanh_distilbert_2020} gives an even slightly better results than \acs{bert}-small~\citep{turc_well-read_2019} while also having fewer parameters than the original \ac{bert}.

For our implementation, the \ac{oss} library transformers\footnote{\url{https://huggingface.co/docs/transformers/index}} from Huggingface~\citep{wolf_huggingfaces_2020} is used.
This provides an abstraction of the actual PyTorch\footnote{\url{https://pytorch.org/}} models and provides easy access to many different pre-trained models.
DistilBERT~\citep{sanh_distilbert_2020}, \acs{bert}-small and \acs{bert}-medium~\citep{turc_well-read_2019}, among others, are available via the transfomers \ac{api}.

\subsubsection{Training}
\label{subsubsec:training}

The entire dataset is composed of descriptions from the \ac{nvd} and texts retrieved from the selected domains~\see{\secref{subsec:implementation_web_scraping}}.
We crawled \num{88979} \ac{nvd} descriptions and \num{12755} texts, for a total of \num{101734} datapoints.
In the following, \ac{nvd} descriptions and retrieved texts are treated identically, \ie, the origin of texts is ignored.

The dataset is split into a training set with 75\% and a test set with 25\% of the texts.
It is ensured that texts referring to the \ac{cve} ID are always also in the same set.

\begin{table}
\centering
\caption{Batch size and training duration of the different models~(Dist. = DistilBERT, Sm. = BERT-small, Med. = BERT-medium) depending on the used \acsp{gpu}.}
\label{tab:batch_size_time}
\begin{tabular}{@{}llrrrrrr@{}}
\toprule
\multicolumn{2}{c}{GPU info} &\multicolumn{3}{c}{Batch size} & \multicolumn{3}{c}{Time [min]} \\
\cmidrule(l{2pt}r{2pt}){1-2} \cmidrule(l{2pt}r{2pt}){3-5} \cmidrule(l{2pt}r{2pt}){6-8} 
Model  & Mem. & Dist. & Sm. & Med. & Dist. & Sm. & Med. \\
A100    & 40Gb     & 48  & 128 & 56 & 60  & 25 & 35 \\
V100    & 32Gb     & 40  & 96  & 48 & 132 & 50 & 95 \\
T40/K80 & 16Gb     & 24  & 64  & 28 & -   & -  & -  \\
\bottomrule
\end{tabular}
\end{table}

The training of the individual DistilBERT, BERT-small, and BERT-medium models is performed independently on \makeanon{the Lichtenberg high-performance computer}.
It provides \acp{gpu} of type Nvidia Ampere 100 and Volta 100.
The batch size is set based on the available \acs{gpu}.
\tabref{tab:batch_size_time} shows the possible batch size and time needed for six epochs of training including evaluation after each epoch.
As one be seen, the training time does not decrease quite linearly with batch size.
The speed of \acsp{gpu} also plays an important role.
In experiments, the training could also be performed on Nvidia T40 and K80 with 16Gb memory.
\citet{shahid_cvss-bert_2021} freeze the layers of the \ac{bert} model for the first three epochs of training and only let the classifier adapt.

\section{Evaluation}
\label{sec:evaluation}

The previously trained models are evaluated in this section.
For this purpose, different metrics for the individual classifiers are considered and compared, including white-box indicators to reconstruct the decision process of our models.
Finally, we determine whether the additional texts have an impact on the overall score.

\subsection{Classifier}
\label{subsec:eval_classifier}

\tabref{tab:metrics} shows various metrics~(Accuracy, Recall, Precision, F1, Cohen $\kappa$) of our classifiers.
The F1 scores are arithmetic means~(macro weighted), so the different distribution of target variables is not taken into account. 

\begin{table*}
\centering
\caption{Accuracy~(Acc), Recall~(Rec), Precision~(Prec), F1 Score, and Cohen's~$\kappa$~(Cohen) for the eight classifiers~(\acf{av}, \acf{ac}, \acf{pr}, \acf{ui}, \acf{s}, \acf{c}, \acf{i}) of each model. Calculated on the test set.}
\label{tab:metrics}
\sisetup{table-format=1.2}
\begin{tabular}{@{} c SSSSS SSSSS SSSSS @{}}
\toprule
\multicolumn{1}{l}{} & \multicolumn{5}{c}{\textbf{DistilBERT}} & \multicolumn{5}{c}{\textbf{BERT-small}} & \multicolumn{5}{c}{\textbf{BERT-medium}} \\
\cmidrule(l{2pt}r{2pt}){2-6} \cmidrule(l{2pt}r{2pt}){7-11} \cmidrule(l{2pt}r{2pt}){12-16} 
 & {Acc} & {Rec} & {Prec} & {F1} & {Cohen} & {Acc} & {Rec} & {Prec} & {F1} & {Cohen} & {Acc} & {Rec} & {Prec} & {F1} & {Cohen} \\ \addlinespace
\acs{av}    & 0.93 & 0.82 & 0.85 & 0.84 & 0.84  & 0.91 & 0.82 & 0.82 & 0.82 & 0.81  & 0.93 & 0.83 & 0.85 & 0.84 & 0.82 \\
\acs{ac}    & 0.96 & 0.78 & 0.85 & 0.82 & 0.64  & 0.96 & 0.8  & 0.84 & 0.83 & 0.61  & 0.94 & 0.79 & 0.84 & 0.8 & 0.61  \\
\acs{pr}    & 0.87 & 0.8 & 0.81 & 0.8 & 0.74    & 0.86 & 0.8  & 0.81 & 0.8  & 0.73  & 0.87 & 0.8 & 0.82 & 0.8 & 0.74   \\
\acs{ui}    & 0.95 & 0.94 & 0.93 & 0.93 & 0.87  & 0.93 & 0.94 & 0.94 & 0.94 & 0.88  & 0.94 & 0.94 & 0.94 & 0.94 & 0.88 \\
\acs{s}     & 0.95 & 0.92 & 0.93 & 0.92 & 0.85  & 0.95 & 0.91 & 0.93 & 0.92 & 0.85  & 0.95 & 0.92 & 0.93 & 0.93 & 0.86 \\
\acs{c}     & 0.89 & 0.85 & 0.86 & 0.87 & 0.8   & 0.89 & 0.85 & 0.86 & 0.87 & 0.8   & 0.88  & 0.85 & 0.87 & 0.86 & 0.8 \\
\acs{i}     & 0.9 & 0.87 & 0.9 & 0.89 & 0.83    & 0.88 & 0.87 & 0.89 & 0.87 & 0.81  & 0.88 & 0.87 & 0.89 & 0.88 & 0.83 \\
\acs{a}     & 0.9 & 0.75 & 0.8 & 0.77 & 0.81    & 0.9  & 0.72 & 0.8  & 0.75 & 0.8   & 0.9 & 0.76 & 0.77 & 0.76 & 0.81  \\
\bottomrule
\end{tabular}
\end{table*}

All models, except the \ac{a} model, achieve F1-scores above \num{0.8} for all components.
The quality of our classifiers is thus comparable to the classifiers of \citet{shahid_cvss-bert_2021}.
However, a clear improvement cannot be seen from the metrics.

For \ac{av}, all models achieve very good predictions for the overrepresented values \textit{N} and \textit{L}.
Although the values \textit{P} and \textit{A} occur very rarely, the classifiers still manage to correctly detect over 70\%.

The classifiers for \ac{ac}, \ac{pr}, and \ac{a} work for frequent values, but are much worse for less frequent ones.
While the F1 score for \ac{ac} and \ac{pr} is unremarkable in each case, this imbalance is evident in the lower Cohen's~$\kappa$.
In particular, for \ac{ac} \textit{H}, the classifiers are not reliable in this way.
For \ac{a}, only around 40\% are correctly detected for \textit{L}, which is the lowest rate of all classifiers.

For \ac{ui}, \ac{s}, \ac{c}, and \ac{i}, the classifiers are good to very good, with the best results for \ac{ui}.

Overall, a highly uneven distribution of values in the dataset tends to lead to worse results in predicting the underrepresented values, which is a common problem with \ac{dl}.

\subsection{\acs*{cvss} Score}
\label{subsec:eval_score}

\begin{table}
\centering
\caption{\acf*{mse}, \acf*{mae}, and proportion of correct~(Pred$_c$), too high~(Pred$_h$), and too low~(Pred$_l$) predictions.}
\label{tab:eval_mse_mae}
\sisetup{table-format=1.3}
\begin{tabular}{@{}l SS S[table-format=2.1]S[table-format=2.1]S[table-format=2.1]@{\hskip 4pt}}
\toprule
                                  & {MSE}  & {MAE} & {Pred$_c$} & {Pred$_h$} & {Pred$_l$} \\
DistilBERT                        & {\cellcolor{green}} 1.44 & {\cellcolor{green}} 0.61 & 62,1\%  & 20,5\% & 17,3\%       \\
BERT-small                        & 1.52                   & 0.624                  & 61,6\%  & 20\%   & 18,2\%       \\
BERT-medium                       & 1.47                   & 0.617                  & 61,6\%  & 20\%   & 18,1\%       \\
\cite{shahid_cvss-bert_2021}      & 1.79                   & 0.73                   & 55,3\%  & {-}    & {-}          \\
\cite{spanos_multi-target_2018}  & {-}                    & 1.74                   & {-}     & {-}    & {-}          \\
\bottomrule
\end{tabular}
\end{table}

To obtain the total \ac{cvss} score, the results of the classifiers of a model are combined.
From the individual components, the score is calculated according to the \ac{cvss} standard\footnote{\url{https://www.first.org/cvss/specification-document}}.
The obtained scores are compared with the expert generated scores in the \ac{nvd}.

\begin{figure}[]
 \centering
 \includegraphics[width=\columnwidth]{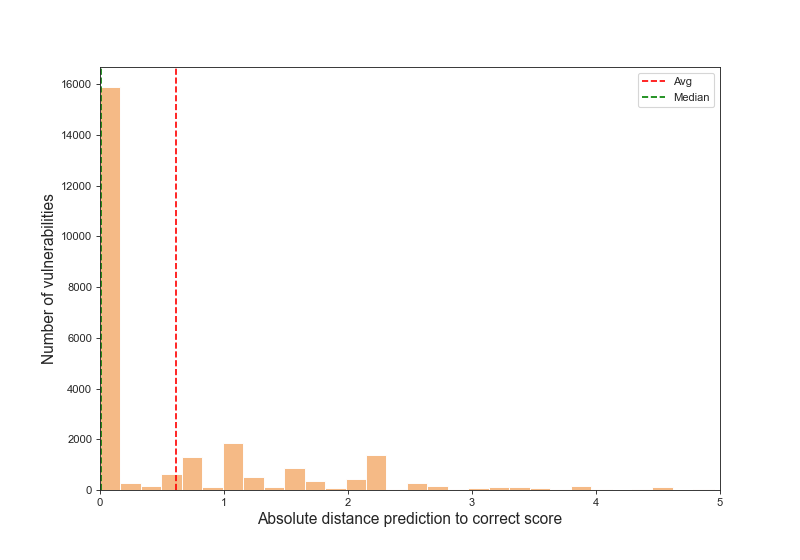}
 \caption{Distribution of difference between the predicted score with the DistilBERT classifiers and \acf*{gt} \acf*{cvss} score.}
 \label{fig:distil_score_diff}
\end{figure}

\tabref{tab:eval_mse_mae} shows the \ac{mae} and \ac{mse} and the fraction of vulnerabilities where the predicted score is higher, lower, or equal to the \ac{nvd}'s \ac{gt}.
It also shows a comparison of our approach against the proposals by \citet{shahid_cvss-bert_2021,spanos_multi-target_2018}.
In \figref{fig:distil_score_diff} the distribution of differences from true to predicted score of DistilBERT classifiers is shown.
The average difference is \num{0.6}, while 75\% of all predictions are within the range of \num{1} around the actual score.

\begin{figure}[]
 \centering
 \includegraphics[width=\columnwidth]{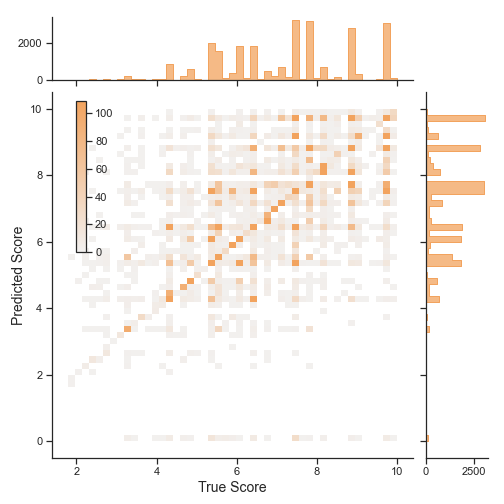}
 \caption{Distribution of DistilBERT classifier predicted scores and the correct \acf*{cvss} scores.}
 \label{fig:hist_distil_scores}
\end{figure}

\figref{fig:hist_distil_scores} shows the ratio of true to predicted scores for the DistilBERT classifiers.
All models predict scores rather higher than lower compared to the original score, but the difference is very small.
In general, it is better to score a vulnerability too high than too low, but depending on the use case, this can be a problem~(\eg, if an information overload is already present).
A striking phenomenon is the series of predictions with a score of \num{0.0}.
These are 111~(DistilBERT), 151~(Bert-small), and 121~(Bert-medium) predictions.
If \textit{N} is predicted for each of the \ac{c}, \ac{i}, and \ac{a} components, this makes the \acf{iss}\footnote{\url{https://www.first.org/cvss/specification-document}} equal to 0.
The \ac{iss} is multiplied by the other \ac{cvss} components, resulting in \num{0.0} for the total score.
There are no vulnerabilities that actually have this combination.
Since there are few such predictions, the problem goes unnoticed in the metrics.
However, critical vulnerabilities~(with scores above 9.0) would be directly discarded because of this problem.

\subsection{Explainability and Interpretability}
\label{subsec:eval_explainability}

For IT security applications, it is important that \ac{ml} procedures are explainable.
\Ac{dl} models usually lack this property.
It is difficult to understand what the model has learned in its entirety.
However, individual examples provide some insight into the model.

\begin{table*}[]
\centering
\caption{Influence of words on prediction of \acf{av}~(with the parameters N - Network, L - Local, A - Adjacent, P - Physical) based on description of \href{https://nvd.nist.gov/vuln/detail/CVE-2016-0775}{CVE-2016-0775}. Green corresponds to a positive influence, red to a negative influence. Predicted \ac{av}: \textit{L}. Correct \ac{av}: \textit{N}}
\label{tab:distil_av_text_expl}
\begin{tabularx}{\textwidth}{cX}
\toprule
\textbf{\ac{av}} & \multicolumn{1}{c}{\textbf{Text}} \\ \addlinespace

N  & \colorbox{red!10}{Buffer} \colorbox{green!10}{over}flow in the ImagingFliDecode function in libImaging/Fl\colorbox{red!10}{iDe}code.c in Pillow before 3.1.1 allows \colorbox{green!30}{remote} \colorbox{green!10}{attackers} to \colorbox{red!10}{cause} a denial of service (crash) via a crafted FLI \colorbox{red!50}{file}. \\ \addlinespace

L  & Buffer overflow in the \colorbox{red!10}{Imaging}FliDecode function in libImaging/\colorbox{red!10}{Fl}\colorbox{green!10}{iDe}\colorbox{green!20}{code}.c in \colorbox{red!10}{Pillow} before 3.1.1 allows remote attackers to cause a denial of service (crash) via a \colorbox{green!10}{crafted} \colorbox{red!10}{FL}I \colorbox{green!50}{file}. \\\addlinespace

A  & Buffer overflow in the ImagingFliDecode function in libImaging/Fl\colorbox{green!10}{iDe}code.c in Pillow before 3.1.1 allows \colorbox{red!40}{remote} \colorbox{red!10}{attackers} to cause a denial of service (crash) via a crafted FLI \colorbox{green!10}{file}. \\\addlinespace

P  & Buffer overflow in the ImagingFliDecode function in libImaging/Fl\colorbox{green!10}{iDe}code.c in Pillow before 3.1.1 allows \colorbox{red!40}{remote} \colorbox{red!10}{attackers} to cause a denial of service (crash) via a crafted FLI \colorbox{green!40}{file}. \\

\bottomrule
\end{tabularx}
\end{table*}

The PyTorch library \textit{Captum}\footnote{\url{https://captum.ai/}} implements various algorithms that help explain \ac{dl} models.
For the following examples, we use Layer Integrated Gradients from Captum on the trained DistilBERT classifiers~\citep{sundararajan_axiomatic_2017}.
\tabref{tab:distil_av_text_expl} shows example description results for the \ac{cvss} prediction of CVE-2016-0775\footnote{\url{https://nvd.nist.gov/vuln/detail/CVE-2016-0775}}.
Words that argue for the classification in each class are marked in green, words that argue against are marked in red.
For clarity, only the words with the greatest influence are marked.
The classifier's prediction is \textit{L}, but \textit{N} would have been correct.

For human experts, the phrase \say{[..]~allows remote attackers to~[..]} is a clear indication that the \acf{av} is \textit{Network}.
As \tabref{tab:distil_av_text_expl} shows, \say{remote attackers} also argues for \textit{N}.
However, the word \say{file} at the end of the description is strongly scored against \textit{N} and for \textit{L}.

This way, a rough understanding of the representation learned by the models is gained.
Many assessments are reasonable, but the \ac{dl} models still remain a black box for users.

\subsection{Influence of Additional Texts}
\label{subsec:eval_influence}

We evaluate whether the texts retrieved via web scrapers have a positive effect compared to using the \ac{nvd}'s descriptions only.
For this purpose, new DistilBERT models are trained on the descriptive texts from \ac{nvd} only.
DistilBERT was chosen because the models performed best overall in our previous evaluation~\see{\tabref{tab:metrics}}.
The models trained exclusively on \ac{nvd} descriptions are called DistilBERT$_\mathit{desc}$ in the following.

\begin{table}
\centering
\caption{\acf*{mse} and \acf*{mae} on different datasets.}
\label{tab:eval_datasets}
\sisetup{table-format=1.3}
\begin{tabular}{@{} l SSSSSS @{}}
\toprule
\multicolumn{1}{c}{} & \multicolumn{2}{c}{Combined} & \multicolumn{2}{c}{Desc} & \multicolumn{2}{c}{Desc$_{2022}$} \\
\cmidrule(l{2pt}r{2pt}){2-3} \cmidrule(l{2pt}r{2pt}){4-5} \cmidrule(l{2pt}r{2pt}){6-7}
                            & {MSE}  & {MAE} & {MSE} & {MAE} & {MSE} & {MAE}      \\
DistilBERT                  & 1.44   & 0.61  & 0.455 & 0.203 & 1.941 & 0.811      \\
DistilBERT$_\mathit{desc}$  & 0.544  & 0.248 & 1.393 & 0.604 & 1.928 & 0.788      \\
\bottomrule
\end{tabular}
\end{table}

\tabref{tab:eval_datasets} shows the results of the evaluation on different datasets.
The \textit{Combined} dataset is the previously used dataset of descriptions~(2016-2021) and retrieved texts.
\textit{Desc} contains only the descriptions, but no additional reference texts.
A new test dataset \textit{Desc$_{2022}$} is used, consisting of \num{5641} descriptions published between January and May 2022.
These descriptions were not previously used for training and evaluation.

On the combined dataset, the DistilBERT$_\mathit{desc}$ classifiers achieve significantly better scores.
Over 80\% of the predictions were correct.
We saw significant improvements for \ac{av}, \ac{ac}, and \ac{a} over the combined trained DistilBERT.
On the \ac{nvd} descriptions, the combined trained DistilBERT is significantly better than DistilBERT$_\mathit{desc}$.
The additional reference texts do have a positive effect.
For \textit{Desc$_{2022}$} the models are on par.
DistilBERT$_\mathit{desc}$ is only slightly better here.

These results are rather unexpected, \ie, the model, which is trained purely on \ac{nvd} descriptions performs significantly better on the prediction of all texts~(including references) and the other way around, and both models perform similar on a new dataset.
The performance of the DistilBERT model trained on the combined dataset can be explained with the higher robustness, but the former cannot.
The only clue might be the high quality of the \ac{nvd} descriptions, but this is counterintuitive to the results of \makeanon{\citet{kuehn_ovana_2021}}.

However, since all models classify texts, which is comparable to predicting a discrete value, small differences might lead to a large difference in the score.
If, for example, the models are tasked to predict the \ac{cvss} score for CVE-2022-23442\footnote{\url{https://nvd.nist.gov/vuln/detail/CVE-2022-23442}, with an expert rated \ac{cvss} vector of \textit{AV:N/AC:L/PR:L/UI:N/S:U/C:L/I:N/A:N}} and it would falsely predict the \ac{c} impact as \textit{None}, the impact score would be \num{0} and with it the whole \ac{cvss} score would result in \num{0}.
There may be a small but crucial difference between the DistilBERT-combined and DistilBERT-descriptions classifiers for a single \ac{cvss} component, but the cause of the surprising results from \tabref{tab:eval_datasets} could not be determined.

\section{Discussion}
\label{sec:discussion}

This section discusses the results~\see{\secref{sec:evaluation}} and points out future work.

\paragraph{Analysis of References and Web Scraping}
The preliminary analysis identified sources of textual vulnerability information besides the \ac{nvd}'s~\rq{1}.
Hereby, we grouped sources and rated their vulnerability uniqueness, uniformity of texts, and the presence of an abstract vulnerability description~\see{\tabref{tab:usability_websites}}.
Due to the strict selection, only references from groups 3, 4 and 5~\see{\secref{subsec:analysis_references}} are eligible.
However, the retrieved texts for this purpose contain almost exclusively the abstract description.
Whether noise would play a major role in the texts is unclear, which could be explored for further work.
While relaxing our criteria would make significantly more web pages usable, the current used language model might not be suited for such task.

Since the used \ac{dl} models are optimized for natural language, log files and source code could not be used.
Some of the references mix code and natural language.
The currently available \ac{nlp} models are not able to use source code in addition to natural language.
Separate models for source code could be used for this in the future.
With such an improvement, future work can build on our reference analysis~\see{\secref{subsec:analysis_references}} and try gather groups with mixed information types.

But, with the current state of research, adaptation to each website is necessary, which increases the effort linearly with the number of websites to create large datasets.
Technically, there is otherwise little potential for optimizations to the implementation of web scraping.
Web scrapers can be parallelized as is and used productively.
Other solutions not based on manually customized web scrapers are not currently available.
The need for manually adapted web scraping would also be eliminated by a uniform standard, \eg,  \ac{csaf}\footnote{\url{https://oasis-open.github.io/csaf-documentation/}}.

\paragraph{Deep Learning Classifier}
Several different current \ac{dl} models were successfully trained and evaluated as classifiers for the components of \ac{cvss} vectors.

The retrieved reference texts could be used as a dataset together with the descriptions.
The obtained classifiers achieve state-of-the-art scores in several metrics~\citep{shahid_cvss-bert_2021, elbaz_fighting_2020}~\see{\tabref{tab:eval_mse_mae}}.
In particular, the DistilBERT model provides good results.
Therefore, the question of whether public data sources beyond databases are suitable for predicting the \ac{cvss} vector~\rq{2} can be answered this way: Texts from \ac{osint} sources are usable for \ac{cvss} prediction, but do not have a clear positive impact on the result in this form.
It is possible to use \ac{osint} as a textual source as a basis for \ac{cvss} prediction.
Since the models require little time to train, it would also be possible to train regularly to incorporate new information into the classifier's decision.
However, the expected positive effect on the quality of the models did not occur due to the additional texts~\see{\secref{sec:evaluation}}.

\paragraph{Limitations \& Future Work}

In the area of web scraping, the paper is limited by the structure of the referenced web pages~\see{\secref{sec:pre-analysis}}.
Future work may simplify web text collection.
This could lower the effort required to adapt web scrapers to different web pages.
Optimally, a solution would be as easy to use as Trafilatura, while still being able to find only the text related to a specific \acs{cve} ID.
Also, a \ac{gui} based program could be developed that allows the selection of elements on a web page.
Based on this selection, the program could then generate the necessary code for the web scraper in a selected web scraping framework.

For the \ac{cvss} classifier it needs to be investigated whether more texts lead to better results.
The influence of noise should be clarified as well and, based on this, the criteria for usable text established in this work should be evaluated again.

The proposed method may have potential for optimization at various points.
Depending on the specific use case, all texts from references could be used for training.
The results in \tabref{tab:eval_datasets} suggest that the additional texts could improve future predictions on new data.
Unrealistic scores with a score of \num{0} can be prevented by minor additions to the logic of the classifiers.
Instead of BERT models trained on general language, models trained specifically on texts from IT security could also serve as a basis.

Data augmentation can be used to improve or compensate for the uneven distribution of different variables in the dataset~\makeanon{\citep{bayer_survey_2022}}. 
\secref{subsec:eval_explainability} has shown that the decisions of the classifiers are only partially understandable.
Further work can improve the explainability and interpretability of the models.

Additionally, the present work lacks a comprehensive comparison against previous work.
This results either from missing published models of previous work to reconstruct the results for a fixed test-set or from the disjoint metrics, that are published.
Hence, a comparison was not possible.

\section{conclusion}
\label{sec:conclusion}

Vulnerabilities in IT systems pose a major threat to society, businesses, and individuals.
A fast and reliable assessment of newly published vulnerabilities is therefore necessary.
The increasing amount of new vulnerabilities makes timely assessment by human experts difficult.
Therefore, various works~\ifreview\citep{shahid_cvss-bert_2021, elbaz_fighting_2020}\else\citep{shahid_cvss-bert_2021,elbaz_fighting_2020,kuehn_ovana_2021}\fi\ deal with automated prediction of \ac{cvss} vector, score or level by \ac{ml}.
These works, however, focused on the \ac{nvd} data alone, rather than using additional \ac{osint} texts for vulnerabilities.
In this work, the possibility of using \ac{osint} as a vulnerability texts source was investigated.
First, a preliminary analysis of the referenced domain in \ac{nvd} vulnerability entries was performed.
In this, the domains are classified into groups based on criteria according to their usability.
This resulted in a pre-selection of domains which later are scraped.
The reference texts and \ac{nvd} descriptions were used as training set for different \ac{dl}-based classifiers.
Finally, the classifiers were evaluated and their quality was assessed.
The classifiers achieve good results in predicting the individual components of the \ac{cvss} vectors.
The \ac{cvss} scores computed from them have low error rates.
However, the \ac{osint} texts have no positive effect on the quality of the classifiers.

\ifreview\else\begin{acks}
We thank all anonymous reviewers of this work.
This research work has been funded by the German Federal Ministry of Education and Research and the Hessian Ministry of Higher Education, Research, Science and the Arts within their joint support of the National Research Center for Applied Cybersecurity ATHENE
and
by the German Federal Ministry for Education and Research~(BMBF) in the project CYWARN~(13N15407).

\end{acks}\fi

\bibliographystyle{ACM-Reference-Format}
\bibliography{bibliography}


\begin{thebibliography}{38}


\ifx \showCODEN    \undefined \def \showCODEN     #1{\unskip}     \fi
\ifx \showDOI      \undefined \def \showDOI       #1{#1}\fi
\ifx \showISBNx    \undefined \def \showISBNx     #1{\unskip}     \fi
\ifx \showISBNxiii \undefined \def \showISBNxiii  #1{\unskip}     \fi
\ifx \showISSN     \undefined \def \showISSN      #1{\unskip}     \fi
\ifx \showLCCN     \undefined \def \showLCCN      #1{\unskip}     \fi
\ifx \shownote     \undefined \def \shownote      #1{#1}          \fi
\ifx \showarticletitle \undefined \def \showarticletitle #1{#1}   \fi
\ifx \showURL      \undefined \def \showURL       {\relax}        \fi
\providecommand\bibfield[2]{#2}
\providecommand\bibinfo[2]{#2}
\providecommand\natexlab[1]{#1}
\providecommand\showeprint[2][]{arXiv:#2}

\bibitem[Almukaynizi et~al\mbox{.}(2017)]%
        {almukaynizi_proactive_2017}
\bibfield{author}{\bibinfo{person}{Mohammed Almukaynizi}, \bibinfo{person}{Eric
  Nunes}, \bibinfo{person}{Krishna Dharaiya}, \bibinfo{person}{Manoj
  Senguttuvan}, \bibinfo{person}{Jana Shakarian}, {and} \bibinfo{person}{Paulo
  Shakarian}.} \bibinfo{year}{2017}\natexlab{}.
\newblock \showarticletitle{Proactive Identification of Exploits in the Wild
  through Vulnerability Mentions Online}. In \bibinfo{booktitle}{\emph{{{CyCon
  U}}.{{S}}. '17}}.
\newblock


\bibitem[Barbaresi(2021)]%
        {barbaresi_trafilatura_2021-1}
\bibfield{author}{\bibinfo{person}{Adrien Barbaresi}.}
  \bibinfo{year}{2021}\natexlab{}.
\newblock \showarticletitle{Trafilatura: {{A Web Scraping Library}} and
  {{Command-Line Tool}} for {{Text Discovery}} and {{Extraction}}}. In
  \bibinfo{booktitle}{\emph{ACL '21: {{System Demonstrations}}}}.
\newblock


\bibitem[Bayer et~al\mbox{.}(2022)]%
        {bayer_survey_2022}
\bibfield{author}{\bibinfo{person}{Markus Bayer}, \bibinfo{person}{Marc-André
  Kaufhold}, {and} \bibinfo{person}{Christian Reuter}.}
  \bibinfo{year}{2022}\natexlab{}.
\newblock \showarticletitle{Survey on {Data} {Augmentation} for {Text}
  {Classification}}.
\newblock \bibinfo{journal}{\emph{{CSUR} '22}} (\bibinfo{year}{2022}).
\newblock


\bibitem[Breiman(2001)]%
        {breiman_random_2001}
\bibfield{author}{\bibinfo{person}{Leo Breiman}.}
  \bibinfo{year}{2001}\natexlab{}.
\newblock \showarticletitle{Random {{Forests}}}.
\newblock \bibinfo{journal}{\emph{Machine Learning '01}}
  (\bibinfo{year}{2001}).
\newblock


\bibitem[{Brian Krebs}(2021)]%
        {brian_krebs_basic_2021}
\bibfield{author}{\bibinfo{person}{{Brian Krebs}}.}
  \bibinfo{year}{2021}\natexlab{}.
\newblock \bibinfo{title}{A {Basic} {Timeline} of the {Exchange} {Mass}-{Hack}
  – {Krebs} on {Security}}.
\newblock
\newblock


\bibitem[Chen et~al\mbox{.}(2019a)]%
        {chen_vase_2019-1}
\bibfield{author}{\bibinfo{person}{Haipeng Chen}, \bibinfo{person}{Jing Liu},
  \bibinfo{person}{Rui Liu}, \bibinfo{person}{Noseong Park}, {and}
  \bibinfo{person}{V.S. Subrahmanian}.} \bibinfo{year}{2019}\natexlab{a}.
\newblock \showarticletitle{{{VASE}}: {{A Twitter-Based Vulnerability
  Analysis}} and {{Score Engine}}}. In \bibinfo{booktitle}{\emph{{ICDM} '19}}.
\newblock


\bibitem[Chen et~al\mbox{.}(2019b)]%
        {chen_vest_2019}
\bibfield{author}{\bibinfo{person}{Haipeng Chen}, \bibinfo{person}{Jing Liu},
  \bibinfo{person}{Rui Liu}, \bibinfo{person}{Noseong Park}, {and}
  \bibinfo{person}{V.~S. Subrahmanian}.} \bibinfo{year}{2019}\natexlab{b}.
\newblock \showarticletitle{{{VEST}}: {{A System}} for {{Vulnerability Exploit
  Scoring}} \& {{Timing}}}. In \bibinfo{booktitle}{\emph{{IJCAI} '19}}.
\newblock


\bibitem[Chen et~al\mbox{.}(2019c)]%
        {chen_using_2019}
\bibfield{author}{\bibinfo{person}{Haipeng Chen}, \bibinfo{person}{Rui Liu},
  \bibinfo{person}{Noseong Park}, {and} \bibinfo{person}{V.S. Subrahmanian}.}
  \bibinfo{year}{2019}\natexlab{c}.
\newblock \showarticletitle{Using {{Twitter}} to {{Predict When
  Vulnerabilities}} Will Be {{Exploited}}}. In \bibinfo{booktitle}{\emph{{KDD}
  '19}}.
\newblock


\bibitem[Devlin et~al\mbox{.}(2019)]%
        {devlin_bert_2019}
\bibfield{author}{\bibinfo{person}{Jacob Devlin}, \bibinfo{person}{Ming-Wei
  Chang}, \bibinfo{person}{Kenton Lee}, {and} \bibinfo{person}{Kristina
  Toutanova}.} \bibinfo{year}{2019}\natexlab{}.
\newblock \bibinfo{title}{{{BERT}}: {{Pre-training}} of {{Deep Bidirectional
  Transformers}} for {{Language Understanding}}}.
\newblock
\newblock


\bibitem[Dong et~al\mbox{.}(2019)]%
        {dong_towards_2019}
\bibfield{author}{\bibinfo{person}{Ying Dong}, \bibinfo{person}{Wenbo Guo},
  \bibinfo{person}{Yueqi Chen}, \bibinfo{person}{Xinyu Xing},
  \bibinfo{person}{Yuqing Zhang}, {and} \bibinfo{person}{Gang Wang}.}
  \bibinfo{year}{2019}\natexlab{}.
\newblock \showarticletitle{Towards the {{Detection}} of {{Inconsistencies}} in
  {{Public Security Vulnerability Reports}}}. In
  \bibinfo{booktitle}{\emph{{{USENIX Security}} '19}}.
\newblock


\bibitem[Dwoskin and Adam(2017)]%
        {dwoskin_nations_2017}
\bibfield{author}{\bibinfo{person}{Elizabeth Dwoskin} {and}
  \bibinfo{person}{Karla Adam}.} \bibinfo{year}{2017}\natexlab{}.
\newblock \showarticletitle{Nations Race to Contain Widespread Hacking}.
\newblock \bibinfo{journal}{\emph{Washington Post}} (\bibinfo{year}{2017}).
\newblock


\bibitem[Elbaz et~al\mbox{.}(2020)]%
        {elbaz_fighting_2020}
\bibfield{author}{\bibinfo{person}{Cl{\'e}ment Elbaz}, \bibinfo{person}{Louis
  Rilling}, {and} \bibinfo{person}{Christine Morin}.}
  \bibinfo{year}{2020}\natexlab{}.
\newblock \showarticletitle{Fighting {{N-day}} Vulnerabilities with Automated
  {{CVSS}} Vector Prediction at Disclosure}. In
  \bibinfo{booktitle}{\emph{{ARES} '20}}.
\newblock


\bibitem[Freund and Schapire(1999)]%
        {freund_short_1999}
\bibfield{author}{\bibinfo{person}{Yoav Freund} {and}
  \bibinfo{person}{Robert~E. Schapire}.} \bibinfo{year}{1999}\natexlab{}.
\newblock \showarticletitle{A {{Short Introduction}} to {{Boosting}}}. In
  \bibinfo{booktitle}{\emph{{JJSAI} '19}}.
\newblock


\bibitem[Gawron et~al\mbox{.}(2018)]%
        {gawron_automatic_2018}
\bibfield{author}{\bibinfo{person}{Marian Gawron}, \bibinfo{person}{Feng
  Cheng}, {and} \bibinfo{person}{Christoph Meinel}.}
  \bibinfo{year}{2018}\natexlab{}.
\newblock \showarticletitle{Automatic {{Vulnerability Classification Using
  Machine Learning}}}. In \bibinfo{booktitle}{\emph{{CRiSIS} '17}}.
\newblock


\bibitem[Gong et~al\mbox{.}(2019)]%
        {gong_joint_2019}
\bibfield{author}{\bibinfo{person}{Xi Gong}, \bibinfo{person}{Zhenchang Xing},
  \bibinfo{person}{Xiaohong Li}, \bibinfo{person}{Zhiyong Feng}, {and}
  \bibinfo{person}{Zhuobing Han}.} \bibinfo{year}{2019}\natexlab{}.
\newblock \showarticletitle{Joint {{Prediction}} of {{Multiple Vulnerability
  Characteristics Through Multi-Task Learning}}}. In
  \bibinfo{booktitle}{\emph{{{ICECCS}} '19}}.
\newblock


\bibitem[Greenberg(2018)]%
        {greenberg_untold_2018}
\bibfield{author}{\bibinfo{person}{Andy Greenberg}.}
  \bibinfo{year}{2018}\natexlab{}.
\newblock \showarticletitle{The {{Untold Story}} of {{NotPetya}}, the {{Most
  Devastating Cyberattack}} in {{History}}}.
\newblock  (\bibinfo{year}{2018}).
\newblock


\bibitem[Han et~al\mbox{.}(2017)]%
        {han_learning_2017}
\bibfield{author}{\bibinfo{person}{Zhuobing Han}, \bibinfo{person}{Xiaohong
  Li}, \bibinfo{person}{Zhenchang Xing}, \bibinfo{person}{Hongtao Liu}, {and}
  \bibinfo{person}{Zhiyong Feng}.} \bibinfo{year}{2017}\natexlab{}.
\newblock \showarticletitle{Learning to {{Predict Severity}} of {{Software
  Vulnerability Using Only Vulnerability Description}}}. In
  \bibinfo{booktitle}{\emph{{{ICSME}} '17}}.
\newblock


\bibitem[Hochreiter and Schmidhuber(1997)]%
        {hochreiter_long_1997}
\bibfield{author}{\bibinfo{person}{Sepp Hochreiter} {and}
  \bibinfo{person}{J{\"u}rgen Schmidhuber}.} \bibinfo{year}{1997}\natexlab{}.
\newblock \showarticletitle{Long {{Short-Term Memory}}}.
\newblock \bibinfo{journal}{\emph{Neural Computation}} (\bibinfo{year}{1997}).
\newblock


\bibitem[Jiang and Atif(2020)]%
        {jiang_approach_2020}
\bibfield{author}{\bibinfo{person}{Yuning Jiang} {and} \bibinfo{person}{Yacine
  Atif}.} \bibinfo{year}{2020}\natexlab{}.
\newblock \showarticletitle{An {{Approach}} to {{Discover}} and {{Assess
  Vulnerability Severity Automatically}} in {{Cyber-Physical Systems}}}.
\newblock In \bibinfo{booktitle}{\emph{{SIN} '20}}.
\newblock


\bibitem[Johnson et~al\mbox{.}(2018)]%
        {johnson_can_2018-1}
\bibfield{author}{\bibinfo{person}{Pontus Johnson}, \bibinfo{person}{Robert
  Lagerstr{\"o}m}, \bibinfo{person}{Mathias Ekstedt}, {and}
  \bibinfo{person}{Ulrik Franke}.} \bibinfo{year}{2018}\natexlab{}.
\newblock \showarticletitle{Can the {{Common Vulnerability Scoring System}} Be
  {{Trusted}}? {{A Bayesian Analysis}}}.
\newblock \bibinfo{journal}{\emph{{TDSC} '18}} (\bibinfo{year}{2018}).
\newblock


\bibitem[Khazaei et~al\mbox{.}(2016)]%
        {khazaei_automatic_2016}
\bibfield{author}{\bibinfo{person}{Atefeh Khazaei}, \bibinfo{person}{Mohammad
  Ghasemzadeh}, {and} \bibinfo{person}{Vali Derhami}.}
  \bibinfo{year}{2016}\natexlab{}.
\newblock \showarticletitle{An Automatic Method for {{CVSS}} Score Prediction
  Using Vulnerabilities Description}.
\newblock \bibinfo{journal}{\emph{Journal of Intelligent \& Fuzzy Systems}}
  (\bibinfo{year}{2016}).
\newblock


\bibitem[Kuehn et~al\mbox{.}(2021)]%
        {kuehn_ovana_2021}
\bibfield{author}{\bibinfo{person}{Philipp Kuehn}, \bibinfo{person}{Markus
  Bayer}, \bibinfo{person}{Marc Wendelborn}, {and} \bibinfo{person}{Christian
  Reuter}.} \bibinfo{year}{2021}\natexlab{}.
\newblock \showarticletitle{{{OVANA}}: {{An Approach}} to {{Analyze}} and
  {{Improve}} the {{Information Quality}} of {{Vulnerability Databases}}}. In
  \bibinfo{booktitle}{\emph{{ARES} '21}}.
\newblock


\bibitem[Le et~al\mbox{.}(2021)]%
        {le_survey_2021}
\bibfield{author}{\bibinfo{person}{Triet H.~M. Le}, \bibinfo{person}{Huaming
  Chen}, {and} \bibinfo{person}{M.~Ali Babar}.}
  \bibinfo{year}{2021}\natexlab{}.
\newblock \showarticletitle{A {{Survey}} on {{Data-driven Software
  Vulnerability Assessment}} and {{Prioritization}}}.
\newblock  (\bibinfo{year}{2021}).
\newblock


\bibitem[Liao et~al\mbox{.}(2016)]%
        {liao_acing_2016}
\bibfield{author}{\bibinfo{person}{Xiaojing Liao}, \bibinfo{person}{Kan Yuan},
  \bibinfo{person}{XiaoFeng Wang}, \bibinfo{person}{Zhou Li},
  \bibinfo{person}{Luyi Xing}, {and} \bibinfo{person}{Raheem Beyah}.}
  \bibinfo{year}{2016}\natexlab{}.
\newblock \showarticletitle{Acing the {{IOC Game}}: {{Toward Automatic
  Discovery}} and {{Analysis}} of {{Open-Source Cyber Threat Intelligence}}}.
  In \bibinfo{booktitle}{\emph{{CCS} '16}}.
\newblock


\bibitem[Liu et~al\mbox{.}(2019)]%
        {liu_vulnerability_2019}
\bibfield{author}{\bibinfo{person}{Kai Liu}, \bibinfo{person}{Yun Zhou},
  \bibinfo{person}{Qingyong Wang}, {and} \bibinfo{person}{Xianqiang Zhu}.}
  \bibinfo{year}{2019}\natexlab{}.
\newblock \showarticletitle{Vulnerability {{Severity Prediction With Deep
  Neural Network}}}. In \bibinfo{booktitle}{\emph{{{BigDIA}} '19}}.
\newblock


\bibitem[Mcauliffe and Blei(2007)]%
        {mcauliffe_supervised_2007}
\bibfield{author}{\bibinfo{person}{Jon Mcauliffe} {and} \bibinfo{person}{David
  Blei}.} \bibinfo{year}{2007}\natexlab{}.
\newblock \showarticletitle{Supervised {{Topic Models}}}. In
  \bibinfo{booktitle}{\emph{{{NIPS}} '07}}.
\newblock


\bibitem[{Pastor-Galindo} et~al\mbox{.}(2020)]%
        {pastor-galindo_not_2020}
\bibfield{author}{\bibinfo{person}{Javier {Pastor-Galindo}},
  \bibinfo{person}{Pantaleone Nespoli}, \bibinfo{person}{F{\'e}lix
  G{\'o}mez~M{\'a}rmol}, {and} \bibinfo{person}{Gregorio
  Mart{\'i}nez~P{\'e}rez}.} \bibinfo{year}{2020}\natexlab{}.
\newblock \showarticletitle{The {{Not Yet Exploited Goldmine}} of {{OSINT}}:
  {{Opportunities}}, {{Open Challenges}} and {{Future Trends}}}.
\newblock \bibinfo{journal}{\emph{IEEE Access}} (\bibinfo{year}{2020}).
\newblock


\bibitem[Ruohonen(2019)]%
        {ruohonen_look_2019}
\bibfield{author}{\bibinfo{person}{Jukka Ruohonen}.}
  \bibinfo{year}{2019}\natexlab{}.
\newblock \showarticletitle{A Look at the Time Delays in {{CVSS}} Vulnerability
  Scoring}.
\newblock \bibinfo{journal}{\emph{Applied Computing and Informatics}}
  (\bibinfo{year}{2019}).
\newblock


\bibitem[Sabottke et~al\mbox{.}(2015)]%
        {sabottke_vulnerability_2015}
\bibfield{author}{\bibinfo{person}{Carl Sabottke}, \bibinfo{person}{Octavian
  Suciu}, {and} \bibinfo{person}{Tudor Dumitraș}.}
  \bibinfo{year}{2015}\natexlab{}.
\newblock \showarticletitle{Vulnerability {{Disclosure}} in the {{Age}} of
  {{Social Media}}: {{Exploiting Twitter}} for {{Predicting Real-World
  Exploits}}}. In \bibinfo{booktitle}{\emph{{USENIX Security} '15}}.
\newblock


\bibitem[Sahin and Tosun(2019)]%
        {sahin_conceptual_2019}
\bibfield{author}{\bibinfo{person}{Sefa~Eren Sahin} {and} \bibinfo{person}{Ayse
  Tosun}.} \bibinfo{year}{2019}\natexlab{}.
\newblock \showarticletitle{A {{Conceptual Replication}} on {{Predicting}} the
  {{Severity}} of {{Software Vulnerabilities}}}. In
  \bibinfo{booktitle}{\emph{{{EASE}} '19}}.
\newblock


\bibitem[Sanh et~al\mbox{.}(2020)]%
        {sanh_distilbert_2020}
\bibfield{author}{\bibinfo{person}{Victor Sanh}, \bibinfo{person}{Lysandre
  Debut}, \bibinfo{person}{Julien Chaumond}, {and} \bibinfo{person}{Thomas
  Wolf}.} \bibinfo{year}{2020}\natexlab{}.
\newblock \bibinfo{title}{{{DistilBERT}}, a Distilled Version of {{BERT}}:
  Smaller, Faster, Cheaper and Lighter}.
\newblock
\newblock


\bibitem[Shahid and Debar(2021)]%
        {shahid_cvss-bert_2021}
\bibfield{author}{\bibinfo{person}{Mustafizur Shahid} {and}
  \bibinfo{person}{Herv{\'e} Debar}.} \bibinfo{year}{2021}\natexlab{}.
\newblock \showarticletitle{{{CVSS-BERT}}: {{Explainable Natural Language
  Processing}} to {{Determine}} the {{Severity}} of a {{Computer Security
  Vulnerability}} from Its {{Description}}}.
\newblock  (\bibinfo{year}{2021}).
\newblock


\bibitem[Spanos and Angelis(2018)]%
        {spanos_multi-target_2018}
\bibfield{author}{\bibinfo{person}{Georgios Spanos} {and}
  \bibinfo{person}{Lefteris Angelis}.} \bibinfo{year}{2018}\natexlab{}.
\newblock \showarticletitle{A Multi-Target Approach to Estimate Software
  Vulnerability Characteristics and Severity Scores}.
\newblock \bibinfo{journal}{\emph{{JSS} '18}} (\bibinfo{year}{2018}).
\newblock


\bibitem[Sundararajan et~al\mbox{.}(2017)]%
        {sundararajan_axiomatic_2017}
\bibfield{author}{\bibinfo{person}{Mukund Sundararajan}, \bibinfo{person}{Ankur
  Taly}, {and} \bibinfo{person}{Qiqi Yan}.} \bibinfo{year}{2017}\natexlab{}.
\newblock \bibinfo{title}{Axiomatic {{Attribution}} for {{Deep Networks}}}.
\newblock
\newblock


\bibitem[Turc et~al\mbox{.}(2019)]%
        {turc_well-read_2019}
\bibfield{author}{\bibinfo{person}{Iulia Turc}, \bibinfo{person}{Ming-Wei
  Chang}, \bibinfo{person}{Kenton Lee}, {and} \bibinfo{person}{Kristina
  Toutanova}.} \bibinfo{year}{2019}\natexlab{}.
\newblock \bibinfo{title}{Well-{{Read Students Learn Better}}: {{On}} the
  {{Importance}} of {{Pre-training Compact Models}}}.
\newblock
\newblock


\bibitem[Wolf et~al\mbox{.}(2020)]%
        {wolf_huggingfaces_2020}
\bibfield{author}{\bibinfo{person}{Thomas Wolf}, \bibinfo{person}{Lysandre
  Debut}, \bibinfo{person}{Victor Sanh}, \bibinfo{person}{Julien Chaumond},
  \bibinfo{person}{Clement Delangue}, \bibinfo{person}{Anthony Moi},
  \bibinfo{person}{Pierric Cistac}, \bibinfo{person}{Tim Rault},
  \bibinfo{person}{R{\'e}mi Louf}, \bibinfo{person}{Morgan Funtowicz},
  \bibinfo{person}{Joe Davison}, \bibinfo{person}{Sam Shleifer},
  \bibinfo{person}{Patrick {von Platen}}, \bibinfo{person}{Clara Ma},
  \bibinfo{person}{Yacine Jernite}, \bibinfo{person}{Julien Plu},
  \bibinfo{person}{Canwen Xu}, \bibinfo{person}{Teven~Le Scao},
  \bibinfo{person}{Sylvain Gugger}, \bibinfo{person}{Mariama Drame},
  \bibinfo{person}{Quentin Lhoest}, {and} \bibinfo{person}{Alexander~M. Rush}.}
  \bibinfo{year}{2020}\natexlab{}.
\newblock \bibinfo{title}{{{HuggingFace}}'s {{Transformers}}:
  {{State-of-the-art Natural Language Processing}}}.
\newblock
\newblock


\bibitem[Yamamoto et~al\mbox{.}(2015)]%
        {yamamoto_text-mining_2015}
\bibfield{author}{\bibinfo{person}{Yasuhiro Yamamoto}, \bibinfo{person}{Daisuke
  Miyamoto}, {and} \bibinfo{person}{Masaya Nakayama}.}
  \bibinfo{year}{2015}\natexlab{}.
\newblock \showarticletitle{Text-{{Mining Approach}} for {{Estimating
  Vulnerability Score}}}. In \bibinfo{booktitle}{\emph{{{BADGERS}} '15}}.
\newblock


\bibitem[Yitagesu et~al\mbox{.}(2021)]%
        {yitagesu_automatic_2021}
\bibfield{author}{\bibinfo{person}{Sofonias Yitagesu},
  \bibinfo{person}{Xiaowang Zhang}, \bibinfo{person}{Zhiyong Feng},
  \bibinfo{person}{Xiaohong Li}, {and} \bibinfo{person}{Zhenchang Xing}.}
  \bibinfo{year}{2021}\natexlab{}.
\newblock \showarticletitle{Automatic {{Part-of-Speech Tagging}} for {{Security
  Vulnerability Descriptions}}}. In \bibinfo{booktitle}{\emph{{{MSR}} '21}}.
\newblock


\end{thebibliography}

\begin{acronym}
\acro{a}[A]{Availability Impact}
\acro{i}[I]{Integrity Impact}
\acro{c}[C]{Confidentiality Impact}
\acro{ac}[AC]{Attack Complexity}
\acro{api}[API]{Application Programming Interface}
\acro{av}[AV]{Attack Vector}
\acro{bert}[BERT]{Bidirectional Encoder Representations from Transformers}
\acro{csaf}[CSAF]{Common Security Advisory Framework}
\acro{cnn}[CNN]{Convolutional Neural Network}
\acro{cnnvd}[CNNVD]{China National Vulnerability Database of Information Security}
\acro{cve}[CVE]{Common Vulnerabilities and Exposures}
\acro{cvss}[CVSS]{Common Vulnerability Scoring System}
\acro{dl}[DL]{Deep Learning}
\acro{dom}[DOM]{Document Object Model}
\acro{gpu}[GPU]{Graphics Processing Unit}
\acro{gt}[GT]{Ground Truth}
\acro{gui}[GUI]{Graphical User Interface}
\acro{http}[HTTP]{Hyper Text Transfer Protocol}
\acro{html}[HTML]{Hypertext Markup Language}
\acro{iaas}[IaaS]{Infrastructure as a Service}
\acro{ioc}[IoC]{Indicators of Compromise}
\acro{iss}[ISS]{Impact Sub Score}
\acro{lstm}[LSTM]{Long Short-Term Memory}
\acro{mae}[MAE]{Mean Absolute Error}
\acro{ml}[ML]{Machine Learning}
\acro{mse}[MSE]{Mean Squared Error}
\acro{nlp}[NLP]{Natural Language Processing}
\acro{nn}[NN]{Neural Network}
\acro{npm}[npm]{Node Package Manager}
\acro{nvd}[NVD]{National Vulnerability Database}
\acro{osint}[OSINT]{Open Source Intelligence}
\acro{oss}[OSS]{Open-Source-Software}
\acro{pos}[POS]{Part-of-Speech}
\acro{pr}[PR]{Privileges Required}
\acro{s}[S]{Scope}
\acro{svm}[SVM]{Support Vector Machine}
\acro{ui}[UI]{User Interaction}
\acro{url}[URL]{Uniform Resource Locator}
\end{acronym}

\end{document}